\documentclass[structabstract]{aa}  
\usepackage{graphicx}
\usepackage{txfonts}
\usepackage{natbib}
\usepackage{color}
\usepackage{lscape}
\bibpunct{(}{)}{;}{a}{}{,} 
\begin{document}
\definecolor{Red}{rgb}{1,0,0}

   \title{The environmental dependence of the stellar mass function at $z\sim1$}
   \subtitle{Comparing cluster and field between the GCLASS and UltraVISTA surveys}
   \author{Remco F.J. van der Burg\inst{1}
          \and Adam Muzzin\inst{1}
          \and Henk Hoekstra\inst{1} 
          \and Chris Lidman\inst{2}
          \and Alessandro Rettura\inst{3}
          \and Gillian Wilson\inst{4}
          \and H.K.C. Yee\inst{5}	
          \and Hendrik Hildebrandt\inst{6,7} 
		  \and Danilo Marchesini\inst{8}
		  \and Mauro Stefanon\inst{9}
		  \and Ricardo Demarco\inst{10}
		  \and Konrad Kuijken\inst{1}
          }
          
   \institute{Leiden Observatory, Leiden University, P.O. Box 9513, 2300 RA Leiden, The Netherlands\\
                 \email{vdburg@strw.leidenuniv.nl}        
             \and Australian Astronomical Observatory, PO Box 915, North Ryde, NSW 1670, Australia
             \and Department of Astronomy, California Institute of Technology, MC 249-17, Pasadena, CA 91125, USA
             \and Department of Physics and Astronomy, University of California-Riverside, 900 University Avenue, Riverside, CA 92521, USA
             \and Department of Astronomy \& Astrophysics, University of Toronto, Toronto, Ontario M5S 3H4, Canada
             \and University of British Columbia, Department of Physics and Astronomy, 6224 Agricultural Road, Vancouver, B.C. V6T 1Z1, Canada
             \and Argelander-Institut f\"ur Astronomie, Auf dem H¨ugel 71, 53121 Bonn, Germany
             \and Tufts University, Robinson Hall, 212 College Avenue, Medford, MA 02155, USA
             \and Department of Physics and Astronomy, Physics Building, University of Missouri, Columbia, MO 65211, USA
             \and Department of Astronomy, Universidad de Concepcion, Casilla 160-C, Concepcion, Chile
             }
   \date{Submitted 4 February 2013; Accepted 30 May 2013}
  \abstract {}
    {We present the stellar mass functions (SMFs) of star-forming and quiescent galaxies from observations of 10 rich, red-sequence selected, clusters in the Gemini Cluster Astrophysics Spectroscopic Survey (GCLASS) in the redshift range $0.86 < z < 1.34$. We compare our results with field measurements at similar redshifts using data from a K$\rm{_s}$-band selected catalogue of the COSMOS/UltraVISTA field.}
    {We construct a K$_{\mathrm{s}}$-band selected multi-colour catalogue for the clusters in 11 photometric bands covering $u$-8$\mu$m, and estimate photometric redshifts and stellar masses using SED fitting techniques. To correct for interlopers in our cluster sample, we use the deep spectroscopic component of GCLASS, which contains spectra for 1282 identified cluster and field galaxies taken with Gemini/GMOS. This allows us to correct cluster number counts from a photometric selection for false positive and false negative identifications. Both the photometric and spectroscopic samples are sufficiently deep that we can probe the stellar mass function down to masses of $10^{10} \,\mathrm{M}_{\odot}$.}
    {We distinguish between star-forming and quiescent galaxies using the rest-frame U-V versus V-J diagram, and find that the best-fitting Schechter parameters $\alpha$ and $M^{*}$ are similar within the uncertainties for these galaxy types within the different environments. However, there is a significant difference in the shape and normalisation of the total stellar mass function between the clusters and the field sample. This difference in the total stellar mass function is primarily a reflection of the increased fraction of quiescent galaxies in high-density environments. We apply a simple quenching model that includes components of mass- and environment-driven quenching, and find that in this picture $45^{+4}_{-3}\%$ of the star-forming galaxies, which normally would be forming stars in the field, are quenched by the cluster.} 
    {If galaxies in clusters and the field quench their star formation via different mechanisms, these processes have to conspire in such a way that the shapes of the quiescent and star-forming SMF remain similar in these different environments.}

   \keywords{Galaxies: clusters: general -- Galaxies: mass function -- Galaxies: evolution -- Galaxies: photometry }
   \maketitle
%

\section{Introduction}
One of the missing parts in the theory of galaxy formation and evolution is a detailed understanding of the build up of stellar mass in the Universe. While the hierarchical growth of dark matter haloes has been studied in large N-body simulations \citep[e.g. ][]{springel05}, the baryonic physics that regulates the cooling of gas and formation of stars in these haloes is much harder to simulate and is not yet well understood. 
To understand which physical processes are dominant in shaping the stellar content of galaxies, models need good observational constraints. 
One of the most fundamental observables of a population of galaxies is their stellar mass function (SMF), which describes the number density of galaxies as a function of stellar mass. Measuring the SMF as a function of cosmic time provides useful constraints on the parameters in semi-analytic models, and these models have to match and predict the SMF for a range of redshifts and environmental densities. 

Although models are tuned to match the observations at $z$=0, there is in general still a poor agreement between observations and theory at higher redshift. Models generally show an excess of galaxies with a stellar mass $(\rm{M_{\star}})\sim 10^{10}\,\rm{M_{\odot}}$ around $z=$1-2 compared to observational data \citep[e.g.][]{bower12,weinmann12}. At higher redshifts the number of high-mass galaxies is generally underpredicted by the models. For a detailed comparison between models and the observed SMF also see \citet{marchesini09}. 

At low redshifts ($z\lesssim 0.2$) the SMF has been measured from wide field data and spectroscopic information \citep{cole01,bell03}, while at higher redshifts the SMF has been measured from deep surveys by making use of photometric redshift estimates \citep{perez08,marchesini09,ilbert10}. The general consensus is that the total stellar mass density evolves slowly between $0<z<1$, which can also be inferred from the sharp decline of cosmic star formation in the Lilly-Madau diagram \citep{lilly96,madau96} in this redshift range. The main evolution is in the normalisation of the SMF, whereas the shape does not show a substantial evolution since $z\sim 4$ \citep{perez08}. However, since these deep surveys generally probe small volumes, the dominant source of random uncertainty is often cosmic variance \citep{somerville04,scoville07,marchesini09}, which is expected to not only have an effect on the normalisation but also on the shape of the observed SMF \citep{trenti08}. Observations over large areas, or a combination of multiple sight lines, are used to reduce this source of uncertainty.

Besides the general time evolution of the properties of galaxies, they are also observed to be strongly influenced by the density of their environment. In particular, galaxies in overdense regions show lower star formation rates, and a higher fraction of red galaxies. At low redshifts, the Sloan Digital Sky Survey has allowed us to quantify these correlations with high precision \citep{kauffmann04,balogh04,blanton05}. The fraction of galaxies that are red is also a function of their stellar mass, with more massive galaxies being redder and forming fewer stars. The quenching fraction of galaxies being a function of both stellar mass and environmental density, some recent studies have suggested the processes of "mass quenching" and "environmental quenching" to be operating completely independently from each other \citep{peng10,muzzin12}, each operating on different time scales. Naively, we would expect the combination of these processes to affect the shape of the SMF.

A measurement of the SMF of galaxies as a function of environmental density therefore provides further constraints on the physical processes that are important in these dense regions. For example, galaxies falling into massive galaxy clusters are expected to be stripped of their cold gas component due to ram-pressure stripping, and a lack of new inflowing cold gas leads to a galaxy's star formation being turned off. Galaxies in groups and clusters are also expected to interact gravitationally through mergers and experience strong tidal forces as they fall towards the cluster centre. 

Combining these measurements done over a range of redshifts and environments puts constraints on the way galaxies quench their star formation, since it allows one to separate between internally and externally driven processes. Some studies have attempted to measure the SMF as a function of local environment at $0.4\lesssim z \lesssim 1.2$ \citep[e.g.][]{bundy06,bolzonella10,vulcani11,vulcani12,giodini12}. A measurement of the SMF at the highest densities has not yet been achieved in this redshift range. This is partly because the deep (and therefore limited in area) surveys that have been used for SMF measurements (mostly the COSMOS and DEEP2 fields) do not contain the extreme overdensities corresponding to the most massive clusters of galaxies. 

In this paper we present a measurement of the SMF of galaxies in 10 rich galaxy clusters at a range of redshifts ($0.86<z<1.34$). These clusters were detected using the red-sequence method on data from the Spitzer Adaptation of the Red-sequence Cluster Survey \citep[SpARCS, see ][]{muzzin09, wilson09, demarco10}, and have typical velocity dispersions of $\sigma_v = 700$ km/s which imply halo masses of $\rm{M_{200}}\simeq 3\times 10^{14}\,\rm{M_{\odot}}$. We combine deep photometric data in 11 bands with the extensive deep spectroscopic coverage that we obtained from the Gemini Cluster Astrophysics Spectroscopic Survey \citep[GCLASS, ][]{muzzin12}. This allows us to estimate stellar masses for individual objects and quantify the amount of interlopers in the photo-$z$ selected sample as a function of mass and projected clustercentric distance. We use the UVJ-diagram to photometrically separate between star-forming and quiescent galaxies, which is critical because the two galaxy types suffer from different observational difficulties and completenesses. We also provide a comparison between the cluster results and the SMF measured from UltraVISTA/COSMOS field. 

The structure of the paper is as follows. In Sect.~\ref{sec:sampledata} we give an overview of GCLASS, and the spectroscopic and photometric data that have been taken for this cluster sample. We also describe the data from the reference UltraVISTA survey. In Sect.~\ref{sec:analysis} we present our measurements of photometric redshifts, stellar masses and rest-frame colours to distinguish between star-forming and quiescent galaxies. We also explain how we correct the photometric sample for incompleteness by making use of the spectroscopic data. In Sect.~\ref{sec:results} we present our results and make comparisons between the two galaxy types, and between cluster environments and the field. In Sect.~\ref{sec:discussion} we discuss our results in the context of galaxy evolutionary processes and in particular quenching in these massive clusters. We summarise and conclude in Sect.~\ref{sec:conclusions}. Extra information considering colour measurements and calibration are presented in the Appendices. There we also compare the UltraVISTA field SMF with the field probed by GCLASS outside the clusters to test for possible systematics. 

All magnitudes we quote are in the AB magnitudes system and we adopt $\Lambda$CDM cosmology with $\rm{\Omega_m=0.3}$, $\rm{\Omega_{\Lambda}=0.7}$ and $\rm{H_0=70\, km\, s^{-1}\,  Mpc^{-1}}$.

\section{Sample \& Data description}\label{sec:sampledata}
\subsection{The GCLASS cluster sample}
The GCLASS cluster sample consists of 10 of the richest clusters in the redshift range $0.86<z<1.34$ selected from the 42 square degree SpARCS survey, see Table~\ref{table:overview}. Clusters in the SpARCS survey were detected using the cluster red-sequence detection method developed by \citet{gladdersyee00}, where the $z'-3.6\mu$m colour was used to sample the $4000\AA$ break at these redshifts \citep[see][]{muzzin08}. 
For an extended description of the SpARCS survey we refer to \citet{muzzin09}, \citet{wilson09} and \citet{demarco10}. 
The 10 clusters that were selected from the SpARCS survey for further study are described in \citet{muzzin12}, and can be considered as a fair representation of IR-selected rich clusters within this redshift range. 

We note that there is a possible selection bias in favour of systems with a high number of bright red galaxies. It is impossible to select clusters based on their total halo mass and therefore any cluster sample has potential selection biases, whether it is X-ray selected, SZ-selected, or galaxy-selected. We note that follow-up studies of X-ray or SZ-selected clusters in the same redshift range also show a significant over-density of red-sequence galaxies \citep[e.g.][]{blakeslee03,mullis05}. Furthermore, the field SMF at $z$=1 shows \citep[e.g.][]{muzzin13b} that even in the field, the bright/massive end of the population is completely dominated by red galaxies. Therefore it seems unlikely that a red-sequence selection results in a significant selection bias, at least for the most massive clusters at a given redshift such as the GCLASS sample.

\begin{table*}[t]
\caption{The 10 GCLASS clusters selected from SpARCS for follow-up spectroscopic and photometric observations. These clusters form the basis of this study.}
\label{table:overview} 
\begin{center}
\centering 
\begin{tabular}{l c r r l l r r} 
\hline\hline 
Name$^{\mathrm{a}}$ & $z_{\mathrm{spec}}$ & RA & Dec &$\rm{K_s}$-band IQ& $\mathrm{K_{lim}}^{\mathrm{b}}$ &$\mathrm{M_{\star,lim}}^{\mathrm{c}}$ &limit from bc03$^{\mathrm{d}}$\\ 
&&J2000&J2000& PSF FWHM ['']&[$\mathrm{mag_{AB}}$]&[$\mathrm{M_{\odot}}$]&\\
\hline 
SpARCS-0034 &0.867 &00:34:42.086 &-43:07:53.360  & 1.01 &21.53 & 10.42&10.43\\
SpARCS-0035 &1.335 &00:35:49.700 &-43:12:24.160  & 0.40 &23.60 & 9.92&9.95\\
SpARCS-0036$^{\mathrm{e}}$ &0.869 &00:36:45.039 &-44:10:49.911  & 1.23(J) &22.11(J) & 10.53&10.50\\
SpARCS-0215 &1.004 &02:15:23.200 &-03:43:34.482  & 1.00 &21.73 & 10.45&10.46\\
SpARCS-1047$^{\mathrm{f}}$ &0.956 &10:47:32.952 & 57:41:24.340  & 0.61 &22.68 & 10.17&10.04\\
SpARCS-1051$^{\mathrm{f}}$ &1.035 &10:51:05.560 & 58:18:15.520  & 0.86 &22.96 & 9.99 &9.99 \\
SpARCS-1613 &0.871 &16:13:14.641 & 56:49:29.504  & 0.81 &22.55 & 9.97&10.02\\
SpARCS-1616 &1.156 &16:16:41.232 & 55:45:25.708  & 0.84 &22.65 & 10.33&10.20\\
SpARCS-1634 &1.177 &16:34:35.402 & 40:21:51.588  & 0.77 &22.88 & 10.14&10.13\\
SpARCS-1638 &1.196 &16:38:51.625 & 40:38:42.893  & 0.66 &23.00 & 10.13&10.09\\
\hline 
\end{tabular}
\end{center}
\begin{list}{}{}
\item[$^{\mathrm{a}}$] For full names we refer to \citet{muzzin12}.
\item[$^{\mathrm{b}}$] 80\% completeness limit for simulated sources. 
\item[$^{\mathrm{c}}$] Corresponding mass completeness limit based on the galaxy in UltraVISTA with the highest M/L fitted at this redshift at $\mathrm{K_{lim}}$. 
\item[$^{\mathrm{d}}$] Mass limit from a synthetic spectrum with $\tau = 10 \rm{Myr}$ starting at age of universe at that redshift with no dust \citep{bc03}.
\item[$^{\mathrm{e}}$] For SpARCS-0036 we used to J-band as the selection band since it is significantly deeper than the $\rm{K_s}$-band. The image quality and magnitude limits refer to the J-band for this cluster.
\item[$^{\mathrm{f}}$] Since the BCG is offset from the centre, this is a better approximation for the cluster centre (different from \citet{muzzin12}). 
\end{list}
\end{table*}

\begin{table}[h]
\caption{Properties of the 10 GCLASS clusters.}
\label{tab:dynanalysis}
\begin{center}
\begin{tabular}{l l l l l}
\hline
\hline
Name & $z_{\mathrm{spec}}$ & $\sigma_{v}$ &$M_{200}$ & $R_{200}$\\
&&$[\rm{km/s}]$&[$10^{14}\,\mathrm{M_{\odot}}$]&[Mpc]\\
\hline
SpARCS-0034&   0.867&$\,\,\,         700_{-         150}^{+          90}$&$\,\,\, 3.6_{- 1.9}^{+ 1.6}$&$ 1.1_{- 0.2}^{+ 0.1}$\\
SpARCS-0035&   1.335&$\,\,\,         780_{-         120}^{+          80}$&$\,\,\, 3.9_{- 1.5}^{+ 1.3}$&$ 0.9_{- 0.1}^{+ 0.1}$\\
SpARCS-0036&   0.869&$\,\,\,         750_{-          90}^{+          80}$&$\,\,\, 4.5_{- 1.4}^{+ 1.6}$&$ 1.1_{- 0.1}^{+ 0.1}$\\
SpARCS-0215&   1.004&$\,\,\,         640_{-         130}^{+         120}$&$\,\,\, 2.6_{- 1.3}^{+ 1.7}$&$ 0.9_{- 0.2}^{+ 0.2}$\\
SpARCS-1047&   0.956&$\,\,\,         660_{-         120}^{+          70}$&$\,\,\, 2.9_{- 1.3}^{+ 1.0}$&$ 1.0_{- 0.2}^{+ 0.1}$\\
SpARCS-1051&   1.035&$\,\,\,         500_{-         100}^{+          40}$&$\,\,\, 1.2_{- 0.6}^{+ 0.3}$&$ 0.7_{- 0.1}^{+ 0.1}$\\
SpARCS-1613&   0.871&$        1350_{-         100}^{+         100}$&$26.1_{- 5.4}^{+ 6.2}$&$ 2.1_{- 0.2}^{+ 0.2}$\\
SpARCS-1616&   1.156&$\,\,\,         680_{-         110}^{+          80}$&$\,\,\, 2.8_{- 1.2}^{+ 1.1}$&$ 0.9_{- 0.1}^{+ 0.1}$\\
SpARCS-1634&   1.177&$\,\,\,         790_{-         110}^{+          60}$&$\,\,\, 4.4_{- 1.6}^{+ 1.1}$&$ 1.0_{- 0.1}^{+ 0.1}$\\
SpARCS-1638&   1.196&$\,\,\,         480_{-         100}^{+          50}$&$\,\,\, 1.0_{- 0.5}^{+ 0.3}$&$ 0.6_{- 0.1}^{+ 0.1}$\\
\hline
\end{tabular}
\end{center}
\end{table}

\subsection{Spectroscopy}\label{sec:spectroscopy}
The clusters in the GCLASS sample have extensive optical spectroscopy, which has been taken using the GMOS instruments on Gemini-North and -South. For details on these measurements, the target selection and an overview of the reduction of these data, we refer to \citet{muzzin12}. 

In summary, spectroscopic targets were selected using a combination of their 3.6$\mu$m fluxes, $z'-3.6\mu$m colours, and their projected clustercentric radii. The colour priority selection was chosen to be sufficiently broad so that there is no selection bias against blue galaxies within the cluster's redshift range. Because the mass-to-light ratio in the 3.6 $\mu$m channel is only a weak function of galaxy type, the targeting completeness is, to first order, a function of radial distance and stellar mass only. The assigned targeting priority is highest for massive objects near the cluster centres \citep[see][Fig.~4]{muzzin12}. 
 
For these 10 clusters there are 1282 galaxies in total with redshifts, of which 457 are cluster members. For more than $90\%$ of the targeted objects with stellar masses exceeding $10^{10}\, \mathrm{M_{\odot}}$, the limiting mass of the photometric data, a redshift was measured with high confidence. Note that the targeting prioritization is known, we do not select against a particular type of galaxies, and we have a high success rate of obtaining redshifts over the stellar mass range we study. Therefore, although the spectroscopic sample is incomplete, it is a representative sample for the underlying population of cluster galaxies. The targeting completeness can be quantified, and in Sect.~\ref{sec:membershipcorrection} we use the spectroscopic sub-sample to correct the full sample for cluster membership.

We have performed a dynamical analysis (Wilson et al., in prep) to study the distribution of line-of-sight (LOS) velocities of the spectroscopic targets. For all 10 clusters, the distribution of LOS velocities approximates a Gaussian profile, which is an indication that the clusters are (close to) virialised. From this distribution we measure the LOS velocity dispersion $(\sigma_{v})$ of the clusters. This leads to estimates of $R_{200}$, the radius at which the mean interior density is 200 times the critical density ($\rho_{\rm{crit}}$), and $M_{200}$, the mass contained within $R_{200}$. The current analysis is done similar to \citet{demarco10}, and is based on an expanded spectroscopic data set. Table~\ref{tab:dynanalysis} shows the cluster properties obtained from this analysis.

The clusters have typical velocity dispersions of $\sigma_v = 700$ km/s which imply halo masses of $\rm{M_{200}}\simeq 3\times 10^{14}\,\rm{M_{\odot}}$. Note that SpARCS-1613 is much more massive, with a velocity dispersion of $\sigma_v = 1350$ km/s. This high value is consistent with the X-ray temperature measured from a recent Chandra observation (see Ellingson, in prep.).

\subsection{Photometric Data}\label{sec:datareduction}
Optical $ugriz$ data for the six clusters observable from the Northern sky were taken with MegaCam at the Canada-France-Hawaii Telescope (CFHT). For the clusters in the South, $ugri$ data were taken with IMACS at the Magellan telescopes, and the $z$-band data using the MOSAIC-II camera mounted on the Blanco telescope at the Cerro Tololo Inter-American Observatory (CTIO). There is J- and $\rm{K_s}$-band imaging data from WIRCam at the CFHT for the Northern clusters, and from ISPI at the Blanco telescope or HAWK-I at the Very Large Telescope (VLT) UT4 for the Southern clusters. Note that these near-IR data were already presented and used in \citet{lidman12} to study the evolution of BCGs. The photometric data set also includes the 3.6, 4.5, 5.8 and 8.0$\mu$m IRAC channels from SWIRE \citep{lonsdale03}. For half of the clusters, including the four at the highest redshifts, we obtained deeper IRAC observations from the GTO programs PID:40033 and PID:50161. The measured depths and an overview of instruments that were used are listed in the Appendix in Table~\ref{table:data}.

In Appendix \ref{catalogcreation} we give a comprehensive description of the photometric data processing leading to a multi-wavelength coverage with a field of view of $10'\times 10'$ centred on the Southern clusters, and a $15'\times 15'$ field of view for the Northern clusters. This wide field view provides information up to several Mpc from the cluster centres at the respective cluster redshifts, even for clusters at the lowest redshifts. 

\subsubsection{Object detection}\label{sec:samplesel}
To measure the stellar mass function it is necessary to obtain a catalogue in which the galaxy sample is complete down to a known mass threshold, independent of their star-formation properties. In an IR-wavelength band the M/L varies little for different star formation histories, so that the luminosity in those bands is a good tracer for the total stellar mass of a galaxy.

Because the IRAC channels suffer from a large point spread function (PSF), separating objects on the sky is difficult. As a compromise between good image quality and detection in a red filter, we therefore choose the $\rm{K_s}$-band as the selection band. We use SExtractor to detect all sources in the $\rm{K_s}$-band that have 5 adjacent pixels with significance $>2.5\sigma$ relative to the local background. 

We obtain a clean catalogue by excluding regions near bright stars and their diffraction spikes, and separate stars from galaxies by using their observed colours. In the $u-\rm{J}$ versus $\rm{J-K}$ colour plane the distinction between stars and galaxies is clear \citep[see e.g.][]{whitaker11}, and we find that the following colour criterion can be used to select a sample of galaxies.
\begin{equation}
 \mathrm{J-K} >0.18\cdot(u- \mathrm{J} )-0.70 \cup \mathrm{J-K} >0.08\cdot(u- \mathrm{J} )-0.40
\end{equation} 

\subsubsection{Colour measurements}
To measure photometric redshifts and stellar masses for the galaxies, accurate colour measurements are necessary. The circumstances of the atmosphere and optical instruments change continuously, and therefore the shape and size of the PSF is different between telescopes, exposures and filters. Therefore it is non-trivial to measure colours of the same intrinsic part of a galaxy. A common approach is to degrade the PSF of the images in all filters to the PSF of the worst seeing, after which the colours are measured by comparing the flux in fixed apertures for all filters. 

An alternative approach, proposed by \citet{kuijken08}, is to perform a convolution of the images in each filter with a position-dependent convolution kernel to make the PSF Gaussian, circular and uniform on each image. The images in the different filters are not required to share the same Gaussian PSF size. Fluxes are measured in apertures with a circular Gaussian weighting function, whose size is adapted to ensure that the same part of the source is measured. 
Because the weighting function approximately matches the galaxy profiles, this technique generally improves the S/N of the measurement compared to a normal top-hat shaped aperture, and we elaborate on this method in Appendix \ref{app:gausspsf}. Note that this technique is not suited for measurements of the total flux, only to obtain colours of a galaxy.

The photometric zeropoints are set based on standard-star observations. We improve the precision of the zeropoints in the $ugriz$JK$_{\mathrm{s}}$ filters by making use of the universality of the stellar locus \citep{slr} and comparing the measured stellar colours in our images with a reference catalogue \citep{covey07}. Further details can be found in Appendix \ref{app:gausspsf}.

\subsection{UltaVISTA field reference}\label{sec:uvista}
In this paper we compare the cluster results to measurements from a representative field at similar redshifts as the clusters. The last decade has seen substantial improvement in the depth and an increase in the field-of-view of ground-based NIR surveys. The most recent wide-field NIR survey is UltraVISTA \citep{mccracken12}, which is composed of deep YJHK$\rm{_s}$ images taken using the VISTA telescope on a 1.6 square degree field that overlaps with COSMOS.

The field sample in this study is based on a K$_{\rm{s}}$-selected catalogue of the COSMOS/UltraVISTA field from \citet{muzzin13a}. The catalogue contains PSF-matched photometry in 30 photometric bands covering the wavelength range 0.15$\mu$m - 24$\mu$m and includes the available $GALEX$ \citep{martin05}, CFHT/Subaru \citep{capak07}, UltraVISTA \citep{mccracken12}, and S-COSMOS \citep{sanders07} datasets.  Sources are selected from the DR1 UltraVISTA K$_{\rm{s}}$-band imaging \citep{mccracken12} which reaches a depth of K$_{\rm{s}} = 23.4$ at 90\% completeness.  A detailed description of the photometric catalogue construction, photometric redshift measurements, and stellar mass estimates is presented in \citet{muzzin13a}. In the next section we estimate these properties for the galaxies selected in GCLASS in a similar way. In Appendix \ref{app:fieldfield} we show a comparison between the UltraVISTA field SMF and the SMF measured in GCLASS outside of the clusters. In general the agreement is good, even though the GCLASS data are much shallower and contain fewer filters. At the low-mass end of the SMF there are some small differences due to incompleteness of GCLASS. We use UltraVISTA to correct this and provide an unbiased measure of the Schechter parameters in the field.

\section{Analysis}\label{sec:analysis}
\subsection{Photometric redshifts}
\begin{figure}
\resizebox{\hsize}{!}{\includegraphics{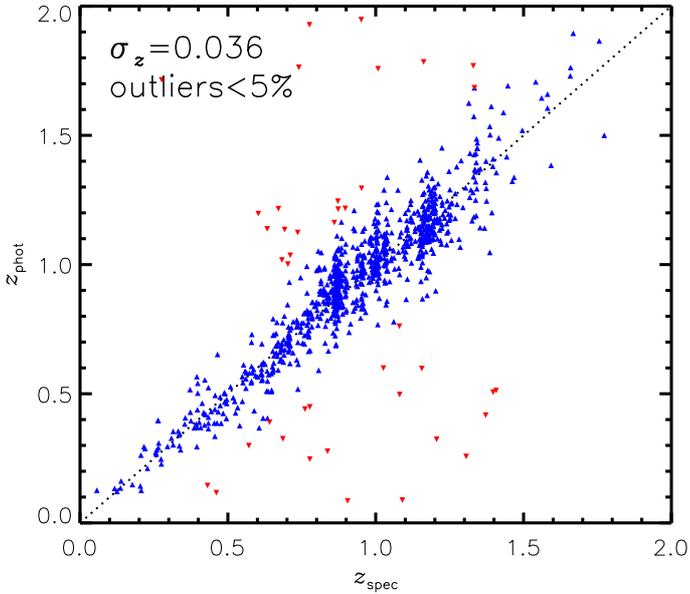}}
\caption{Spectroscopic versus photometric redshifts for the 10 GCLASS clusters. Outliers, objects for which $\Delta z > 0.15$, are marked in red. The outlier fraction is less than $5\%$, the scatter of the remaining objects is $\sigma_z = 0.036$.}
\label{fig:speczphotz_test}
\end{figure}
We estimate photometric redshifts (photo-$z$'s) using the publicly available code EAZY \citep{brammer08}. This code was tested \citep{hildebrandt10} and performs very well on simulated and real imaging data. Input to the code are fluxes in the 11 available bands and their errors. 

We checked for possible systematic problems in the photometric calibration or photo-$z$ code by comparing the estimated photometric redshifts with spectroscopic redshifts where the samples overlap, see Fig.~\ref{fig:speczphotz_test}. The performance is then quantified by the scatter, bias and outlier fraction of this comparison. First we calculate $\Delta z = \frac{z_{\mathrm{phot}}-z_{\mathrm{spec}}}{1+z_{\mathrm{spec}}}$ for each object with a reliable spectroscopic redshift. For historical reasons and to facilitate comparisons with other photo-$z$ studies, we define outliers as objects for which $|\Delta z |> 0.15$. For the remaining measurements we measure the mean of $|\Delta z |$ and the scatter around this mean, $\sigma_z$. We find the following typical values: a scatter of $\sigma_z = 0.036$, a bias of $|\Delta z |= 0.005$, and fewer than $5 \%$ outliers. 
Specifically, in the redshift range of the clusters ($0.867<z<1.335$), we find a scatter of $\sigma_z = 0.035$, a bias of $|\Delta z |< 0.005$, and about $8 \%$ outliers. We find that the scatter is in the range $0.031<\sigma_{z}<0.044$ for all 10 clusters, and therefore the differences in photo-$z$ quality between the clusters is negligible.

Whereas these values are computed for the entire population of galaxies, a subdivision by galaxy type shows that photo-$z$ estimates for quiescent galaxies are more precise ($\sigma_z = 0.030$) than for star-forming galaxies ($\sigma_z = 0.040$) because of the stronger 4000$\AA$ feature in the broad-band SEDs of quiescent galaxies, and the presence of emission lines and a larger diversity of intrinsic SEDs in the star-forming population. We therefore make the separation by galaxy type when correcting for cluster membership in Sect.~\ref{sec:membershipcorrection}. The scatter in photo-$z$ estimates increases for fainter objects, however we take this effect into account when we correct for cluster membership.

\subsection{Stellar masses and completeness}\label{sec:stellarmasses}
We estimate stellar masses for all objects using the SED fitting code FAST \citep{kriek09}. The redshifts are fixed at the measured spec-$z$, whenever available. Otherwise we use the photo-$z$ from EAZY, and the stellar population libraries from \citet{bc03} are used to obtain the model SED that gives the best fit to the photometric data. We use a parameterization of the star formation history as $SFR \propto e^{-t/\tau}$, where the time-scale $\tau$ is allowed to range between 10 Myr and 10 Gyr. We also assume a \citet{chabrier03} IMF, solar metallicity, and the \citet{calzetti00} dust law. These settings are identical to those used for the measurement of stellar masses in the UltraVISTA sample, in order to provide a fair comparison. 

We estimate the mass completeness limits for each of the clusters in the following way. First we measure the depths of the $\rm{K_s}$ detection bands by performing simulations in which we add artificial sources to these images for a range of magnitudes. We then run SExtractor with the same settings as for the construction of the catalogue (Sect.~\ref{sec:samplesel}). The recovered fraction as a function of magnitude for the simulated sources provides an estimate for the depth of the detection image. 
Note that the clusters at higher redshift were prioritized to have longer exposure times and therefore deeper detection bands, leading to more homogeneous detection limits in terms of absolute magnitude and stellar mass. Magnitude values corresponding to the $80\%$ recovery limit, which are typically $\sim \rm{22.5 mag_{\rm{AB}}}$, are given in Table~\ref{table:overview}. 

We estimate stellar mass limits that correspond to these $80\%$ $\rm{K_s}$-band completeness limits in two different ways. The first method uses the UltraVISTA catalogue, which is about a magnitude deeper than GCLASS in the $\rm{K_s}$-band. For each cluster we select all galaxies from UltraVISTA with a photometric redshift within 0.05 from the cluster redshift. By comparing the total $\rm{K_s}$-band magnitudes with estimated stellar masses in this redshift range, we identify the galaxy with the highest mass around the limiting detection magnitude. This is the object with the highest mass-to-light ratio, corresponding to the reddest galaxy in UltraVISTA. All galaxies more massive than these mass limits, which are listed in Table~\ref{table:overview}, will be detected with a probability of $>80\%$ in GCLASS.

Secondly, to provide a comparison, we also give the mass limit corresponding to a maximally old stellar population with no dust \citep{bc03}, at the redshift of the cluster with a flux equal to the detection limit. The mass limits resulting from this approach are also given in Table~\ref{table:overview}, and are similar to the first estimates to within several hundredths of a dex for most of the clusters. 

Note that for cluster SpARCS-0036 we use the J-band as the detection band instead of the $\rm{K_s}$-band because the $\rm{K_s}$-band is of much lower quality. Because the seeing in the J-band is significantly better, a J-band selection leads to a stellar mass detection limit that is 0.3 dex lower than could be obtained with a $\rm{K_s}$-selection. In Table~\ref{table:overview} we therefore give the estimates corresponding to the J-band.

\subsection{Rest-frame colours}\label{sec:restframecols}
In the following we make a separate comparison between the SMF for star-forming and quiescent galaxies, and correct each of the galaxy types for cluster membership. \citet{wuyts07}, \citet{williams09} and \citet{patel12} have shown that the \mbox{U-,} V- and J-band rest-frame fluxes of galaxies can be combined into a UVJ diagram to distinguish quiescent galaxies from star-forming galaxies, even if the latter population is reddened by dust extinction. 

\begin{figure}
\resizebox{\hsize}{!}{\includegraphics{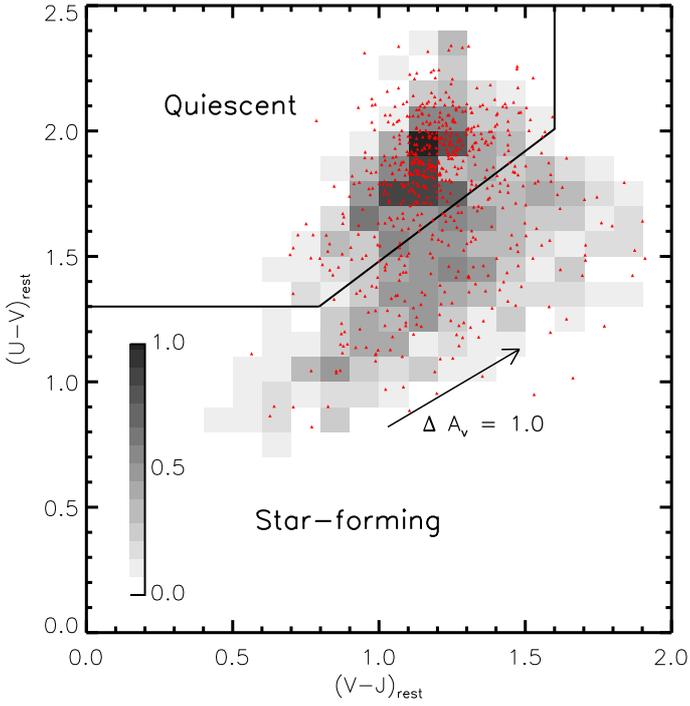}}
\caption{Rest-frame U-V versus V-J colours for galaxies with stellar masses exceeding $10^{10}\, \rm{M_\odot}$ to differentiate between quiescent and star-forming galaxies. The arrow indicates the reddening vector for dust. Combining both colours facilitates a distinction between both galaxy types, even when there is significant reddening by dust. In grey is the distribution of galaxies from GCLASS that are between $0.85<z<1.20$, but outside the clusters. A relative density scale is provided. The red points show photo-$z$ selected cluster members with projected position less than 2 arcmin from the cluster centres.}
\label{fig:UVJ_diagram}
\end{figure}

After estimating redshifts for all objects in the photometric catalogue, we use EAZY to interpolate the input SED to obtain the U-V and V-J rest-frame colours for each galaxy. In Fig.~\ref{fig:UVJ_diagram} we plot those colours for all $\rm{K_s}$-band selected objects with $\rm{M_{\star}} > 10^{10}\, \rm{M_\odot}$. The greyscale distribution shows the galaxies in GCLASS that are in the redshift range $0.85<z<1.20$, but are not part of the clusters, while the red points show the objects that are separated from the BCG by less than 2 arcmin, and have a photometric redshift within 0.1 from the cluster redshift. We select as the quiescent population galaxies with $\mathrm{(U-V) > 1.3 \wedge (V-J) < 1.6 \wedge (U-V) > 0.88(V-J)+0.6}$ \citep[e.g.][]{whitaker11}. This dividing line is shown in the figure. For reference, the dust-reddening vector is also shown, indicative of a dust-independent separation of quiescent and star-forming galaxies. 

Comparing the cluster and field galaxy populations, we find that $68\%$ of the cluster galaxies in this mass range are quiescent, whereas only $42\%$ of the field galaxies are quiescent. This shows that the cluster population is dominated by quiescent galaxies, whereas the field has a more mixed population of galaxies. Note, however, that the distribution of colours due to dust-reddening within the separate galaxy types is similar for the two environments. 

\subsection{Cluster member selection}\label{sec:membershipcorrection}
\begin{figure}
\resizebox{\hsize}{!}{\includegraphics{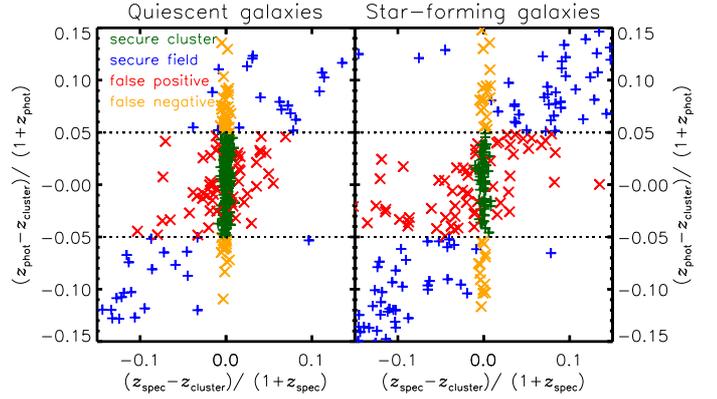}}
\caption{An adaptation of Fig.~\ref{fig:speczphotz_test}, showing a composite plot of the 10 clusters to measure the fraction of false positives and false negatives, after separating quiescent and star-forming galaxies. By plotting the difference with respect to the cluster redshift, all clusters are effectively plotted on top of each other. The $z_{\rm{phot}}$ measurements for star-forming galaxies have a larger scatter than the measurements for quiescent galaxies. What is not shown here, is how the purity fractions change as a function of mass and radial distance. In the analysis we also take account of this mass and radial dependence; see Fig.~\ref{fig:corrfactors_all}.}
\label{fig:completenessbasis}
\end{figure}

Due to the scatter in the photometric redshift estimates, selecting cluster galaxies based on photometric redshifts will result in the sample being contaminated by fore- and background galaxies. In this section we combine the photometric $\rm{K_s}$-band selected multi-colour catalogue and the sub-sample of galaxies with spectroscopic information to select a complete sample of cluster members. We will use the following terminology. By "false positive" we refer to an object that is not part of the cluster, yet has a photo-$z$ that is consistent with the cluster redshift. A "false negative" is an object that belongs to the cluster, but has a photo-$z$ inconsistent with cluster membership. A "secure cluster" object is correctly classified as being part of the cluster based on the photo-$z$, while a "secure field" object is correctly identified as being outside of the cluster.

Given the relatively small fields in which we measure the cluster SMF, field-to-field variance complicates a full statistical interloper subtraction that is based solely on photometric data. However, owing to the extended spectroscopic coverage of GCLASS, we can estimate the field contamination from these data without having to rely on the statistical subtraction of an external field. This way we take account of both false positives and false negatives in the photometrically selected sample. The objects in the spectroscopic sample were prioritized by 3.6$\mu$m IRAC flux and proximity to the cluster core, see Sect.~\ref{sec:spectroscopy} and \citet{muzzin12}. This selection leads to a targeting completeness that is, to first order, a function of radial distance and stellar mass only. 

For these targets we measure the differences between photo-$z$'s and the redshift for each cluster, and between spec-$z$'s and the redshift of the cluster. A composite for all 10 clusters is shown in Fig.~\ref{fig:completenessbasis}, after separating between quiescent and star-forming galaxies. This can be considered as a different representation of Fig.~\ref{fig:speczphotz_test}, where the data for all clusters have been folded on top of each other. Galaxies with $|\Delta z| < 0.05$ are selected as preliminary cluster members based on their photometric redshifts. The red crosses show false positives, orange crosses indicate false negatives. Green (blue) symbols show objects that are identified as secure cluster (field) galaxies. Note that, although we could have started with any cut on $|\Delta z|$, the $|\Delta z| < 0.05$ criterion conveniently yields a number of false positives that approximately equals that of false negatives.

\begin{figure*}[ht]
\resizebox{\hsize}{!}{\includegraphics{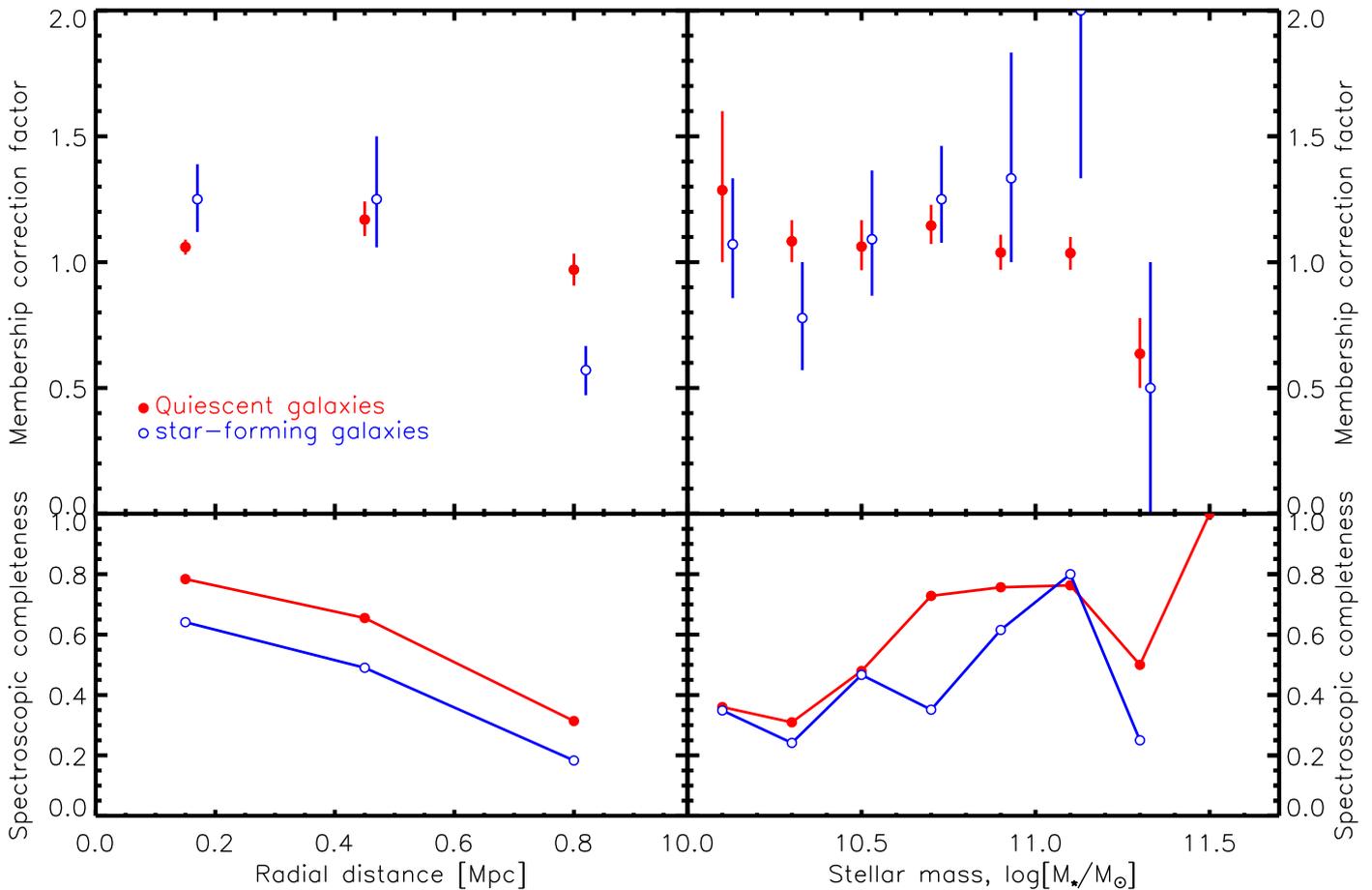}}
\caption{The correction factors for the photo-$z$ selected members that have no spec-$z$ information, estimated from the subsample of objects that do have spectroscopic overlap. A separation by radial distance and stellar mass is made, and these factors are multiplied to yield the total correction factor for each galaxy. A correction factor $>$ 1 indicates that the number of false negatives exceeds the number of false positives in that bin. In the bottom panels the spectroscopic targeting completeness is shown.}
\label{fig:corrfactors_all}
\end{figure*}

For the objects in the photometric sample that do not have a spectroscopic redshift, we use these fractions of false positives and false negatives to correct the number counts for cluster membership. To make sure that the spec-$z$ subsample is representative of the photo-$z$ sample, we have to estimate this correction as a function of radial distance and stellar mass. This separation ensures that we take account of the spectroscopic targeting completeness, the mass- and radially-dependent overdensity of the cluster compared to the field, and the flux dependence of the photo-$z$ quality. In Fig.~\ref{fig:corrfactors_all} we show the correction factors, as a function of radial distance (left panel) and as a function of stellar mass (right panel). Error bars give the 68$\%$ confidence regions estimated from a series of Monte-Carlo simulations in which we randomly draw a number for secure cluster members, false positives and false negatives from a Poisson distribution in each mass-, and radial bin. A correction factor $>$ 1 indicates that the number of false negatives exceeds the number of false positives in that bin. Corrections are roughly constant with $\rm{M_{\star}}$, decreasing slightly at large radii, but the selection of photo-$z$ members as objects with $|\Delta z| < 0.05$ ensures that the corrections are small in general. If we change the cut to 0.03, 0.07 or 0.10, this leads to different membership corrections. However, after these corrections have been applied, we find that these cuts give results that are consistent within the errors. 

Down to the mass-completeness of the clusters there are 283 spectroscopically confirmed cluster members. We divide the 255 photo-$z$ members for which we do not have spectra over a 2-dimensional array of 3 radial bins and 8 stellar mass bins, and correct them for membership by multiplying with both the radial and mass-dependent correction factors (as shown in Fig.~\ref{fig:corrfactors_all}). Because the corrections are relatively small, the way we bin the data only has a minor effect on the results. 
The dominant source of uncertainty is of statistical nature.

\section{Results}\label{sec:results}
\begin{figure*}
\resizebox{\hsize}{!}{\includegraphics{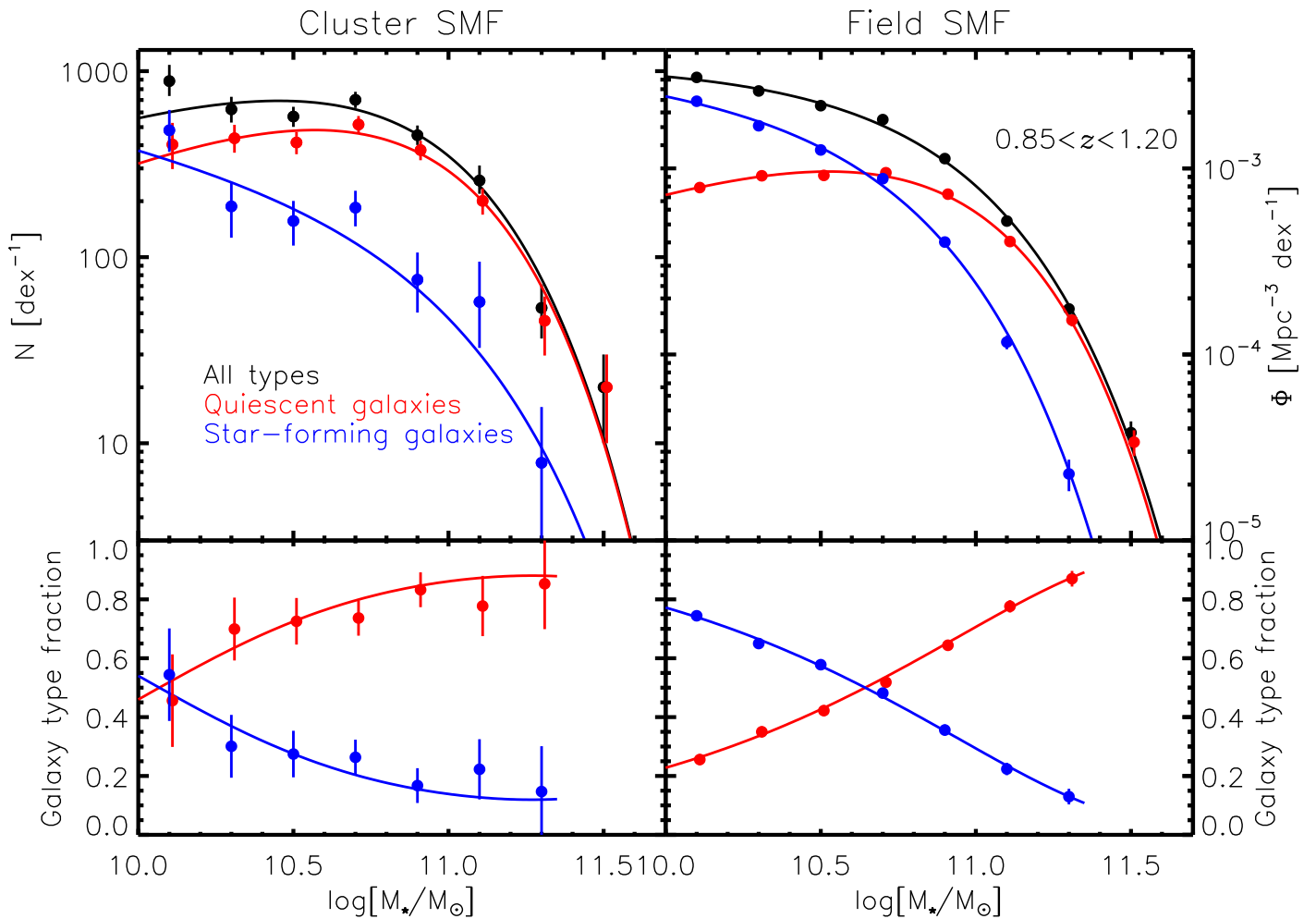}}
\caption{Comparing the cluster SMF (left panel) with a similar representation of the field SMF (right panel). The total SMFs (black points) are separated by galaxy type. Red points show the quiescent galaxies and the blue points show star-forming galaxies. The best-fitting Schechter functions are overplotted for each SMF sample. Note that the red points have been offset by 0.01 dex for better visibility. In the bottom panel we show the fractional contribution of quiescent and star-forming galaxies to the total population, and the curves show the fractional contributions of the Schechter functions. The relative contribution of quiescent galaxies is shown to be higher in the cluster than in the field. Note that the error bars on the field data are smaller than the data point symbols, because only Poissonian errors are taken into account.}
\label{fig:smf_field_and_cluster}
\end{figure*}

\begin{table*} 
\caption{The values for the data points of the galaxy SMF that are shown in Fig.~\ref{fig:smf_field_and_cluster}. These are the raw, membership-corrected, numbers of galaxies for the clusters. To obtain the units shown in the figures for the clusters, these values need to be multiplied by 5, since the binsize is 0.2 dex in stellar mass. Numbers in brackets show the total number of spectroscopic cluster members in each bin. Note that the spectroscopic completeness is highest in the high-mass bins. Errors represent $1\sigma$ uncertainties estimated from Monte-Carlo simulations for the cluster data, and Poissonian errors for the field data.}
\label{table:datapoints} 
\begin{center}
\centering 
\begin{tabular}{c | l l l| l l l } 
\hline\hline 
$\rm{log(}\rm{M_{\star}})$&Cluster $z\sim 1$&Number&&Field $0.85<z<1.20$&$\Phi\, [10^{-5}\, \rm{dex^{-1}\, Mpc^{-3}}]$&\\
$[\rm{M_{\odot}}]$&Total&Quiescent&Star-forming&Total&Quiescent&Star-forming\\
\hline 
 10.10&$   176_{-    29}^{+    39}\,[          24]$&$\,\,\,    80_{-    21}^{+    24}\,[           9]$&$\,\,\,    96_{-    22}^{+    27}\,[          15]$&$308.6\pm  5.1$&$ 78.9\pm  2.6$&$229.7\pm  4.4$\\
 10.30&$   124_{-    18}^{+    20}\,[          20]$&$\,\,\,    87_{-    14}^{+    15}\,[          13]$&$\,\,\,    37_{-    12}^{+    12}\,[           7]$&$260.8\pm  4.7$&$ 91.3\pm  2.8$&$169.5\pm  3.8$\\
 10.50&$   114_{-    13}^{+    14}\,[          46]$&$\,\,\,    82_{-    11}^{+    11}\,[          34]$&$\,\,\,    31_{-     8}^{+     8}\,[          12]$&$217.4\pm  4.3$&$ 91.7\pm  2.8$&$125.7\pm  3.3$\\
 10.70&$   140_{-    13}^{+    14}\,[          78]$&$   103_{-    10}^{+    11}\,[          63]$&$\,\,\,    36_{-     7}^{+     8}\,[          15]$&$183.0\pm  3.9$&$ 94.9\pm  2.8$&$\,\,\, 88.1\pm  2.7$\\
 10.90&$\,\,\,    90_{-    10}^{+    11}\,[          63]$&$\,\,\,    75_{-     8}^{+     9}\,[          55]$&$\,\,\,    15_{-     5}^{+     6}\,[           8]$&$112.9\pm  3.1$&$ 72.7\pm  2.5$&$\,\,\, 40.2\pm  1.
8$\\
 11.10&$\,\,\,    51_{-     7}^{+    10}\,[          33]$&$\,\,\,    40_{-     6}^{+     6}\,[          29]$&$\,\,\,    11_{-     4}^{+     7}\,[           4]$&$\,\,\, 52.1\pm  2.1$&$ 40.5\pm  1.8$&$\,\,\, 11.7\pm  1.0$\\
 11.30&$\,\,\,    10_{-     3}^{+     3}\,[           8]$&$\,\,\,\,\,\,     9_{-     3}^{+     3}\,[           7]$&$\,\,\,\,\,\,     1_{-     1}^{+     1}\,[           1]$&$\,\,\, 17.6\pm  1.2$&$ 15.3\pm  1.1$&$
\,\,\,\,\,\,  2.3\pm  0.4$\\
 11.50&$\,\,\,\,\,\,     4_{-     2}^{+     2}\,[           4]$&$\,\,\,\,\,\,     4_{-     2}^{+     2}\,[           4]$&\,\,\,\,\,\,\,\,\,\,\,\,\,[           0]&$\,\,\,\,\,\,  3.8\pm  0.6$&$\,\,\,  3.4\pm  0.5$
&$\,\,\,\,\,\,  0.4\pm  0.2$\\
   \hline
\end{tabular}
\end{center}
\end{table*}

\subsection{The cluster stellar mass function}\label{sec:clustersmf}
We measure the cluster galaxy SMF from the sample of galaxies in the 10 GCLASS clusters, obtained as described in Sect.~\ref{sec:membershipcorrection}. This is done by summing over the 3 radial bins so that we measure the SMF out to a projected radius of 1 Mpc. The summation is done separately for quiescent and star-forming galaxies, which were identified using the UVJ criterion (Sect.~\ref{sec:restframecols}). The errors from the Monte-Carlo simulations that we discussed in Sect.~\ref{sec:membershipcorrection} are propagated. Note however that the spectroscopic targets only contribute a Poissonian error, since these do not need to be statistically corrected for cluster membership.

The blue points in the left panel of Fig.~\ref{fig:smf_field_and_cluster} show the SMF for the star-forming galaxies in the 10 clusters, while the red points show the quiescent population in the clusters. The total galaxy SMF is the sum of the two galaxy types, and is shown in black. The fraction of quiescent and star-forming galaxies to the total number of galaxies is shown in the bottom panel. The data points are also given in Table~\ref{table:datapoints}. Note that the quiescent population dominates the SMF of the cluster galaxies over almost the entire mass range we study. The BCGs are not included in this plot, nor in the rest of the analysis in this paper. Although the satellites in the galaxy clusters are believed to originate from an infalling population of centrals in the field, the BCGs are the central galaxies in massive cluster haloes and do not have a field counterpart. Consequently, BCGs do not follow the Schechter function that describes the rest of the cluster galaxies. For a study of the stellar mass evolution of BCGs we refer to \citet{lidman12}.

Because the selection bands of some of the clusters are not sufficiently deep to probe the SMF down to $10^{10} \,\mathrm{M_{\odot}}$ (see Table~\ref{table:overview}), the lowest two stellar-mass bins are composed of galaxies selected from 6 and 7 clusters, respectively. These two bins were scaled up by assuming the richnesses of the clusters are similar, i.e. multiplying their values with a factor of $\frac{10}{6}$ and $\frac{10}{7}$, respectively. A rough estimate of the richnesses of the individual galaxy clusters shows that these corrections factors are accurate to within 10\%. 

We perform a small additional completeness correction based on a comparison of the field SMF measured from UltraVISTA and the parts of GCLASS that are outside the clusters (i.e. the field SMF from GCLASS; see Appendix \ref{app:fieldfield}). Because of the depth of its photometry, UltraVISTA is complete at $\rm{M_{\star}} > 10^{10}\, \rm{M_\odot}$ in the redshift range $0.85<z<1.20$. We compare the field estimates in Appendix \ref{app:fieldfield} and find that there is a good quantitative agreement in both the shape and normalisation of the field SMF between the surveys at this stellar mass range, except for the lowest three mass points. This suggests that there may be residual incompleteness in GCLASS that affects the lowest mass points. Assuming that this incompleteness affects the cluster and field data of GCLASS in a similar way, we correct the GCLASS cluster SMF points for the star-forming and quiescent galaxies with small factors, up to $37\%$ at the lowest mass bin for the quiescent galaxies. This correction changes the best-fit Schechter parameters for the cluster fits in the following way. $M^{*}$ increases by 0.01, 0.10 and 0.08 dex and $\alpha$ becomes more negative by 0.07, 0.33 and 0.26 for the total, star-forming and quiescent population, respectively. These changes do not affect any of the qualitative results in this paper, nor change the conclusion in any way.

We fit a Schechter \citep{schechter76} function to the binned data points. This function is parameterized as  
\begin{equation}
\Phi(M)= \ln (10) \Phi^{*} \left[ 10^{(M-M^{*})(1+\alpha)}\right]\cdot \exp \left[ -10^{(M-M^{*})}\right],
\end{equation}
with $M^{*}$ being the characteristic mass, $\alpha$ the low-mass slope, and $\phi^{*}$ the total overall normalisation. Our data cannot rule out a different functional form at the low-mass end. Therefore we will discuss the differences in the SMFs between the cluster and field in the context of the Schechter function fit. A more quantitative assessment would require measurements at lower masses.

Because the number of sources in the brighter stellar mass bins is low, we are in the regime where errors cannot be represented by a Gaussian distribution and therefore ordinary $\chi ^2$ minimisation is not appropriate. Consequently, we take a general maximum likelihood approach where we use the probability functions on each data point obtained from the Monte-Carlo simulations. This way we compute the likelihood function $\mathcal{L}$ on a 3 dimensional grid of Schechter parameters. The best fitting parameters $M^{*}$ and $\alpha$, corresponding to $\mathcal{L_{\rm{max}}}$, are listed in Table~\ref{tab:sparms} and the corresponding Schechter function is shown in the left panel of Fig.~\ref{fig:smf_field_and_cluster} (black curve). The Schechter function provides a reasonable fit to the data, with a Goodness of Fit (GoF) of 2.12. We also give the 68.3\% confidence levels on each parameter after marginalising over the other two parameters. We take this confidence interval to be the region where $2 \ln(\mathcal{L_{\rm{max}}}/\mathcal{L}) \leq 1$. However, since these parameters are known to be degenerate, we show confidence regions in Fig.~\ref{fig:total_ellipses}. The black curves correspond to 68.3\% and 95.4\% confidence levels after marginalising only over $\phi^{*}$.

In general, uncertainties on individual mass measurements of the galaxies lead to a bias in the shape of the SMF and the best fitting Schechter parameters \citep{eddington1913,teerikorpi04}. Especially for high masses, where the slope of the SMF is steep, the shape of the SMF can be biased because of galaxies scattering to adjacent bins. To study the magnitude of this effect on our analysis, we need to estimate the stellar mass scatter of galaxies in each bin of the SMF. To do this we created 100 Monte Carlo realisations of the photometric catalogue, in which we randomly perturb the aperture fluxes following the estimated statistical errors on these measurements. Then we estimate photo-$z$'s and stellar masses for the entries of these catalogues in a similar way as for the standard analysis. At the high-mass end, where the SED fitting is mostly supported by spec-$z$'s (see Fig.~\ref{fig:corrfactors_all} or Table~\ref{table:datapoints}), the scatter is about 0.05 dex in stellar mass. For lower masses the scatter increases towards 0.08 (0.10) dex in stellar mass for quiescent (star-forming) galaxies. Even if all galaxies would scatter to higher masses, the bias in the Schechter parameter $M^{*}$ would be 0.05 dex. In reality $\alpha$ might also change slightly due to Eddington bias \citep[e.g.][]{vanderburg10}, but we expect the bias of the combination of Schechter parameters to be substantially smaller than the size of the 1-$\sigma$ statistical error contours in Fig.~\ref{fig:total_ellipses}. Given also that the systematic uncertainties due to assumptions regarding the IMF, models on the stellar populations, star-formation histories and metallicity, are substantially larger \citep[e.g.][]{marchesini09}, we do not attempt to correct for this bias in the current analysis.

\begin{table}
\caption{A comparison between the best fitting Schechter parameters and their 68\% confidence intervals for the different galaxy types and environments.}
\label{tab:sparms}
\begin{center}
\begin{tabular}{ccccc}
\hline\hline
Galaxy type&Environment &log[$M^{*}/\rm{M_{\odot}}$]&$\alpha$&$GoF^{\mathrm{a}}$\\
\hline
Total&Cluster&$10.72^{+0.09}_{-0.02}$&$-0.46^{+0.08}_{-0.26}$&$2.12$ \\
Total&Field&$10.83^{+0.01}_{-0.02}$&$-1.01^{+0.04}_{-0.02}$&$4.66$ \\
Star-forming&Cluster&$10.87^{+0.28}_{-0.18}$&$-1.38^{+0.38}_{-0.35}$&$1.44$ \\
Star-forming&Field&$10.65^{+0.02}_{-0.01}$&$-1.13^{+0.02}_{-0.05}$&$4.15$ \\
Quiescent&Cluster&$10.71^{+0.04}_{-0.10}$&$-0.28^{+0.33}_{-0.14}$&$1.21$ \\
Quiescent&Field&$10.77^{+0.01}_{-0.01}$&$-0.43^{+0.02}_{-0.04}$&$1.39$ \\
\hline
\end{tabular}
\end{center}
\begin{list}{}{}
\item[$^{\mathrm{a}}$] Goodness of Fit (GoF) defined as $\chi^2/\rm{dof}$ for the field survey (we assumed Gaussian statistics owing to the large number counts in this survey). For the cluster fits we used an analogous expression from the Maximum likelihood fitting method.
\end{list}
\end{table}

\begin{figure*}
\resizebox{\hsize}{!}{\includegraphics{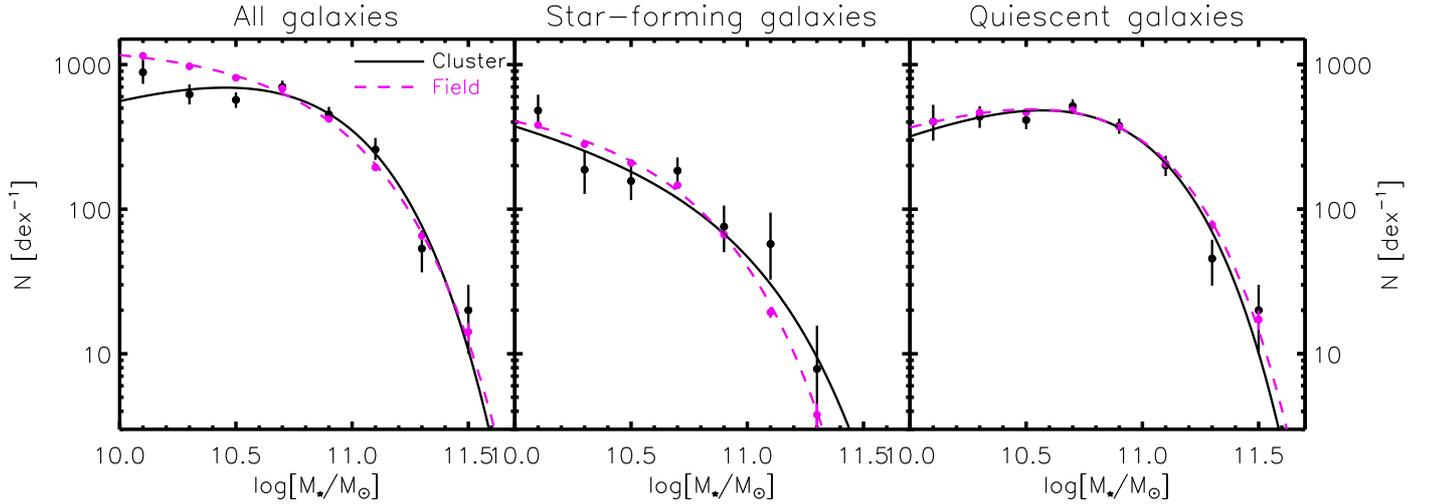}}
\caption{Galaxy SMFs for different galaxy types and environments. Left panel: the total galaxy population in the cluster (black) and the field (magenta). Middle panel: the cluster and field SMF for the subset of star-forming galaxies. Right panel: the subset of quiescent galaxies. The field data have been scaled vertically to match the cluster SMF at $M^*$ of the cluster. Error bars show the $68\%$ confidence regions from Monte-Carlo simulations (on the cluster data), or Poisson error bars (field data).}
\label{fig:smf_superplot}
\end{figure*}

\begin{figure*}
\resizebox{\hsize}{!}{\includegraphics{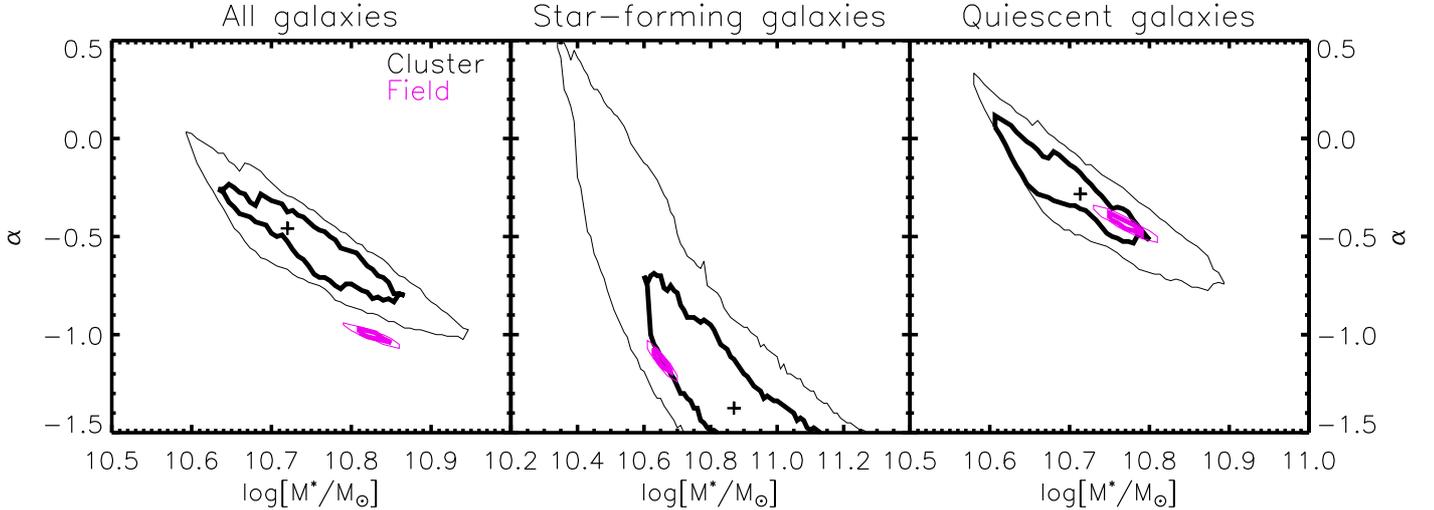}}
\caption{The 68\% and 95\% likelihood contours for the Schechter parameters $M^{*}$ and $\alpha$, after marginalising over the $\phi^*$ parameter. Black lines show the cluster contours, while magenta lines show the contours for the field data. $+$-signs show the single best fit Schechter parameters. The regions corresponding to the cluster SMF were obtained using maximum-likelihood fitting of the Monte-Carlo simulated data.}
\label{fig:total_ellipses}
\end{figure*}

\subsection{Cluster versus Field}
We compare the cluster results with the field galaxy SMF by selecting all galaxies with a photometric redshift in the range $0.85<z<1.20$ from the UltraVISTA survey. Since the UltraVISTA survey is superior in depth compared to GCLASS, the SMF can be measured down to $10^{10}\,\rm{M_{\odot}}$ in this redshift range. The right panel of Fig.~\ref{fig:smf_field_and_cluster} shows the field total SMF in black, which is composed of 13633 galaxies in this mass and redshift range. The best fitting Schechter function for the field sample is found by minimizing $\chi ^2$ on a 3 dimensional grid of Schechter parameters, and is represented by the black curve in the right panel of Fig.~\ref{fig:smf_field_and_cluster}. For a comprehensive comparison between the SMF from UltraVISTA and other field estimates we refer to \citet{muzzin13b}. There it is shown that the SMF of the entire galaxy population, measured with this catalogue, is in good agreement with previous measurements.

To better compare the shape of the total SMF in the two environments, we refer to the left panel of Fig.~\ref{fig:smf_superplot}, where the magenta points show the galaxy SMF from UltraVISTA, and the black points show the SMF for the cluster galaxies.
The field data have been scaled such that the Schechter functions of the cluster and field intersect at the characteristic mass $M^{*}$ of the cluster. The best fitting values for the $\alpha$ and $M^{*}$ parameters are given in Table~\ref{tab:sparms}, with their 68.3\% confidence levels. Because we only included Poissonian errors on the field SMF data, the GoF of the Schechter fits are rather high (up to 4.66 for the total galaxy population). At this level of detail it is also possible that the Schechter function is no longer an adequate description of the data. The magenta contours in the left panel of Fig.~\ref{fig:total_ellipses} show the 2-d confidence contours for the field.

\subsection{Star-forming vs Quiescent galaxies}
We separate the UltraVISTA galaxy catalogue between quiescent galaxies and star-forming galaxies by using their estimated rest-frame U-V and V-J colours, as was analogously done for the cluster galaxies in Sect.~\ref{sec:restframecols}. We compare the shapes of the SMF for each galaxy type between the different environments. 

In the middle panel of Fig.~\ref{fig:smf_superplot} we show the shape of the SMF for star-forming galaxies in the field (magenta) and the cluster (black), together with their best-fitting Schechter functions. The field data have been normalised so that the Schechter functions intersect at the characteristic mass $M^{*}$ for star-forming galaxies in the cluster. The corresponding 68$\%$ and 95$\%$ confidence regions for the Schechter parameters $\alpha$ and $M^{*}$ are shown in the middle panel of Fig.~\ref{fig:total_ellipses}. The best-fitting Schechter parameters and their error bars are also given in Table~\ref{tab:sparms}. 

\subsection{Normalisation of the SMF}
The data points in Fig.~\ref{fig:smf_superplot} are arbitrarily normalised to provide for an easy comparison of the shapes of the SMF between the field and cluster samples. As a consequence, the $\phi^{*}$ parameters corresponding to the best fitting Schechter function have no direct meaning. Normalised by volume the cluster is, by definition, substantially overdense compared to the field. To be able to better interpret the differences of the SMF between the field and cluster environment in Sect.~\ref{sec:discussion}, we therefore normalise the SMF by the total amount of matter in each respective part of the Universe. 

For the UltraVISTA field reference we take the total comoving volume within a redshift range $0.85<z<1.20$ and an unmasked survey area of 1.62 square degree \citep{muzzin13a}. After multiplying the volume corresponding to this area in this redshift range, $5.9\cdot10^{6}\,\rm{Mpc}^{3}$, by the average matter density of the Universe, being 2.8 $\cdot 10^{-30 }\,\rm{g}\,\rm{cm}^{-3}$ in our cosmology, we find that the total amount (i.e. dark matter + baryonic) of matter in this volume is about $2.4\cdot 10^{17}\,\rm{M_{\odot}}$.

Given the values for $M_{200}$, which are presented in Table~\ref{tab:dynanalysis} and are based on the dynamical analysis of the GCLASS spectra, we estimate the concentration parameter corresponding to the NFW profiles \citep{nfw} for these systems from \citet{duffy08}. We integrate these NFW profiles along the LOS and out to a projected radius of 1 Mpc, yielding a total mass of $5.6\times 10^{15} \,\rm{M_{\odot}}$ for the 10 clusters. Since \citet{sheldon09} and \citet{hoekstra00} have shown that, although cluster centres are dominated by luminous matter, the mass to light ratio (M/L) of clusters within a distance of 1 Mpc is similar to the cluster M/L within larger distances, this ensures that we measure and normalise the SMF in a representative volume.

\begin{figure}[hbt]
\resizebox{\hsize}{!}{\includegraphics{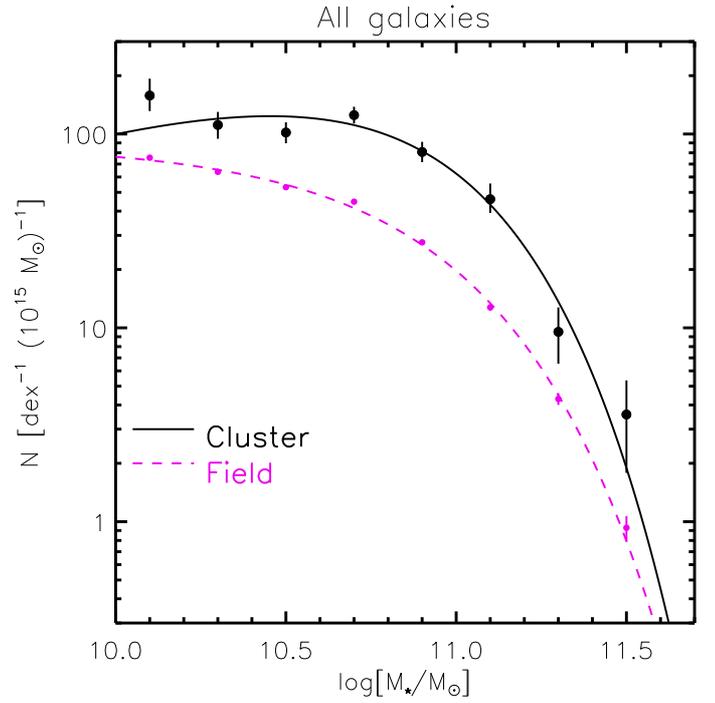}}
\caption{Same as the left panel of Fig.~\ref{fig:smf_superplot}, but normalised by the total mass (dark matter + baryonic) in the field sample (magenta) and cluster sample (black). Per unit of total mass the cluster has a clear overdensity at all stellar masses we probe. Error bars show the $68\%$ confidence regions from Monte-Carlo simulations (on the cluster data), or Poisson error bars (field data). In the left panel of Fig.~\ref{fig:smf_superplot} we provided an easier comparison of the shapes of the two SMFs.}
\label{fig:smf_normalised}
\end{figure}

Fig.~\ref{fig:smf_normalised} shows the total SMF for the cluster and the field, after normalising by the total masses estimated above. Note that there is, per unit total mass, a strong overdensity of galaxies at all stellar masses we probe in the cluster environments. In the stellar mass range we study, the overdensity ranges from a minimum factor of 1.3 at $10^{10} \,\rm{M_{\odot}}$ to a maximum factor of 3.2 at $10^{11.1} \,\rm{M_{\odot}}$. This shows that the clusters contain a very biased population of galaxies, where a relatively high fraction of the total baryonic mass is transformed into stars. The field, in contrast, contains regions such as voids, where the star formation efficiency is very low. 

\section{Discussion}\label{sec:discussion}
In this section we discuss the implications of the results from Sect.~\ref{sec:results}. We discuss in Sect.~\ref{sec:shapesmf} the shape of the SMF for star-forming galaxies, quiescent galaxies, and the total galaxy population. We make a comparison between the cluster and field, and also compare our results to measurements from the literature. We proceed to apply a simple model that \citet{peng10} showed to give a good fit to the SMF measured at $z=0$ from SDSS data. \citet{peng10} could not explore the area of high-$z$ clusters with COSMOS and SDSS data, so we confront our results at $z=1$ with the predictions of their model.

\subsection{The shape of the galaxy SMF}\label{sec:shapesmf}
\subsubsection{Star-forming galaxies}
Fig.~\ref{fig:smf_superplot} shows that the shape of the galaxy SMF for the subset of UVJ-selected star-forming galaxies is similar between the clusters from GCLASS and the field from UltraVISTA. Quantitatively, Fig.~\ref{fig:total_ellipses} indicates that the combination of best-fitting Schechter parameters differs by about $1\sigma$. The low-mass slope $\alpha$ is $-1.38^{+0.38}_{-0.35}$ for the cluster versus $-1.13^{+0.02}_{-0.05}$ in the field. The characteristic mass $M^{*}$ is $10.87^{+0.28}_{-0.18}$ and $10.65^{+0.02}_{-0.01}$ for the cluster and field, respectively. 

We do not make a quantitative comparison between the literature and our measurements of the SMF for star-forming galaxies because the way these star-forming samples are selected is different for different studies. Whereas we select a subset of star-forming galaxies based on the UVJ-diagram, most other studies use either a single colour or a morphological selection. Nonetheless, the finding that the shape of the star-forming SMF is independent of environment is qualitatively consistent with lower redshift measurements presented by e.g. \citet{bolzonella10}. Note however that the clusters in GCLASS constitute much higher overdensities than the highest densities in the COSMOS fields used by \citet{bolzonella10}. The shape of the star-forming galaxy SMF is also measured to be roughly constant with cosmic time \citep[e.g.][]{ilbert10,brammer11}. This shows that, whatever processes are responsible for the quenching of star formation in galaxies, they have to operate in such a way that the SMF of star-forming galaxies does not change shape, even in the highest density environments. This is a fundamental assumption for the \citet{peng10} quenching model that we employ in Sect.~\ref{sec:model}.

\subsubsection{Quiescent galaxies}
Fig.~\ref{fig:smf_superplot} shows that for the selection of quiescent galaxies based on the UVJ criterion, the shape of the SMF for those galaxies is also similar in the different environments probed by GCLASS and UltraVISTA. The best fitting $\alpha$ for the clusters is $-0.28^{+0.33}_{-0.14}$ versus $-0.43^{+0.02}_{-0.04}$ in the field. Given the degeneracy between $\alpha$ and $M^{*}$, the combination of these Schechter parameters, as shown in Fig.~\ref{fig:total_ellipses}, also agrees to better than $1\sigma$ between the field and cluster. It seems remarkable that, whatever quenching processes are responsible for the build-up of the quiescent population in these contrasting environments, they work in such a way that the resulting SMF for quiescent galaxies at $\rm{M_\star > 10^{10}\, M_{\odot}}$ has a similar shape in both environments. 

\citet{rudnick09} measured the cluster galaxy luminosity function of red sequence galaxies in the redshift range $0.4<z<0.8$ and compared their measurements with the field luminosity function. They also found little difference in the shape of the quiescent luminosity function between the two environments. \citet{rudnick09} also found a hint of a shallower low-mass slope in the cluster compared to the field. 
Note that they use a different selection of red galaxies, so that their red sequence selected sample might be contaminated by reddened star-forming galaxies. 

\subsubsection{The total galaxy population}
Whereas the SMF for each of the galaxy types appears to be similar in the different environments probed by GCLASS and UltraVISTA, Figs.~\ref{fig:smf_superplot} \& \ref{fig:total_ellipses} show that the SMF for the total galaxy population is significantly different. This is because the \textit{fraction} of quiescent galaxies is higher in the cluster. That makes the low-mass slope of the total SMF shallower in the cluster compared to the field (see Fig.~\ref{fig:smf_superplot}). This result is also consistent with the measurements shown for more moderate overdensities in the COSMOS field by \citet{bolzonella10}. We compare our results to the literature results from the WINGS, ICBS and EDisCS clusters probed in \citet{vulcani13}, although our sample is unique in this combination of redshift range and photometric depth.

\citet{vulcani13} assumed a Kroupa \citep{kroupa01} IMF, which yields stellar masses consistent with Chabrier to within several 0.01 dex. For the sample of WINGS clusters ($0.04<z<0.07$) they measure Schechter parameters $M^{*}=10.82 \pm 0.13$ and $\alpha = -0.88 \pm 0.31$. Although the redshift distribution is very different from the GCLASS sample, they agree within $1-\sigma$ with the contours shown in Fig.~\ref{fig:total_ellipses}. The measured Schechter parameters for the ICBS clusters ($0.3<z<0.45$) are $M^{*}=11.37 \pm 0.28$ and $\alpha = -1.29 \pm 0.41$. Note that this point lies in the direction of the correlation between $M^{*}$ and $\alpha$, as is shown in Fig.\ref{fig:total_ellipses}. The same is true for the EDisCS clusters ($0.4<z<0.8$), for which \citet{vulcani13} report Schechter parameters $M^{*}=11.15 \pm 0.07$ and $\alpha = -1.03 \pm 0.08$.

Another fundamental observable of a population of galaxies, besides their SMF, is the distribution of specific star formation rates (sSFRs). \citet{wetzel12} studied the distribution of sSFRs for central and satellite galaxies as a function of stellar mass in a range of environments. They show that the distribution of sSFRs is clearly bimodal, with clear populations of active and passive galaxies. Interestingly, they show that the location and shape for each of the two peaks is independent of environment, and that only the relative amounts of star-forming and quiescent galaxies occupying the peaks differ as a function of environment. Likewise, \citet{muzzin12} show that for the GCLASS data the sSFR of star-forming galaxies in a given mass bin is also independent of environment. These results are analogous to our measurements for the SMF, which can also be considered as a sum of the quiescent galaxy SMF and the star-forming galaxy SMF. Having a different fraction of quiescent galaxies in opposing environments, the total galaxy SMF will look different whereas the SMF for each galaxy type is similar, analogous to what \citet{wetzel12} found for the distribution of sSFRs. 

\subsection{A simple quenching model}\label{sec:model}
It has been known for several decades that the fraction of quiescent galaxies increases with both stellar mass and environmental density \citep[e.g.][]{baldry06}. However, recent studies \citep[e.g.][]{peng10,muzzin12} have suggested that the quenching of star-forming galaxies can be fully separated in two distinct quenching tracks, dubbed "mass quenching" and "environmental quenching". The assumption that the shape of the galaxy SMF for star-forming galaxies is universal, which is supported by our measurements, places constraints on the way these quenching processes operate. 

To interpret our observed data in this context we consider the simple model proposed by \citet{peng10}. 
This model is based on the observed constancy in the shape of the SMF for star-forming galaxies with redshift. \citet{peng10} use a combination of mass quenching and environmental quenching, processes which they assume to act independently of each other, to build up the quiescent population. The basic descriptions for these quenching tracks are demanded to operate such that the shape of the SMF for star-forming galaxies is independent of environment. 

Because star-forming galaxies are forming stars at a rate that scales roughly linearly with their stellar mass (the observed sSFR for this population is roughly independent of mass \citep{noeske07}), mass quenching is supposed to preferentially quench high mass galaxies in order to keep the SMF for star-forming galaxies unchanged. Therefore the resulting galaxy SMF for this quenched population is expected to contain an excess of high mass galaxies and hence has a shallow low-mass slope. In high-density environments the fraction of quiescent galaxies increases compared to low-density environments. \citet{peng10} assume that this increase is caused by the process of environmental quenching. The environmental quenching efficiency is assumed to be independent of mass, so that the resulting SMF of the environmentally-quenched galaxies has the same shape as the star-forming galaxy SMF. With some additional quenching due to, what they presume to be, merger processes, \citet{peng10} showed that this model works very well at reproducing the SMF measured in the redshift range $0.02<z<0.085$ from SDSS DR7 \citep{abazajian09} data. The regime of $z\sim 1$ clusters however has not yet been tested against their model, since this range is not probed by COSMOS.

\begin{figure*}
\resizebox{\hsize}{!}{\includegraphics{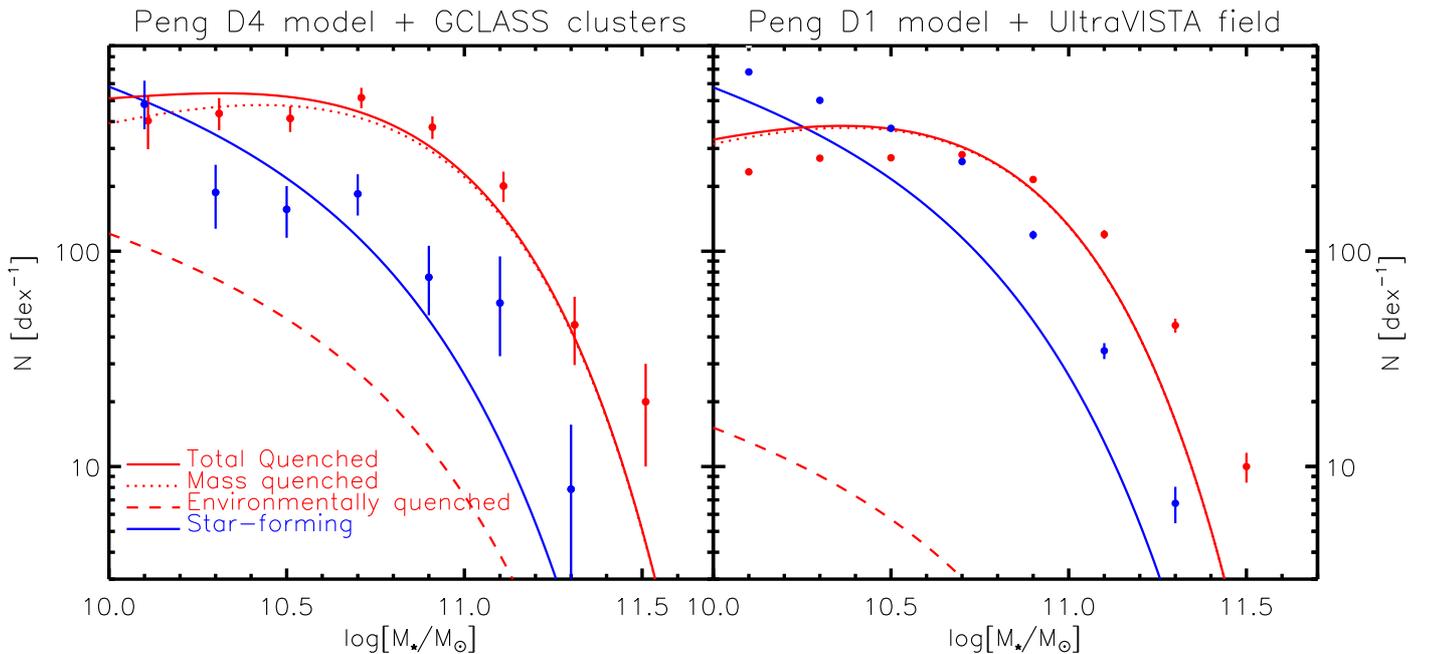}}
\caption{The left panel compares the \citet{peng10} model prediction in the environmental density quartile D4 with our GCLASS cluster SMF measurements, which were already presented in Fig.~\ref{fig:smf_field_and_cluster} and Table~\ref{table:datapoints}. A separation between the two quenching processes is made. The right panel makes a similar comparison between the model in quartile D1 with the UltraVISTA field data. Note that the relative normalisations of star-forming and quiescent populations are fixed, and that the populations are fitted simultaneously.}
\label{fig:smf_peng_model}
\end{figure*}

The model, however, makes predictions for the SMF at higher redshifts over a range of environmental densities \citep[][Fig. 13]{peng10}, and we compare these predictions at $z=1$ to the SMFs measured from GCLASS and UltraVISTA. The predictions from their model are separated by environmental density in four quartiles, with D1 (D4) corresponding to the lowest (highest) density quartile. The (especially environmentally) quenched galaxies contribute more substantially to the total galaxy population in D4 compared to D1, leading to a higher fraction of quiescent galaxies. The left panel of Fig. \ref{fig:smf_peng_model} compares the prediction of the highest environmental density quartile (D4) with the measurement of the cluster galaxy SMF from GCLASS. We fitted the total normalisation as a free parameter, but left the relative normalisations of star-forming and quiescent galaxies unchanged. Note that the GCLASS clusters constitute the most massive structures around $z=1$ and therefore contain higher overdensities than the D4 reference. Nevertheless, the D4 model provides a reasonable fit to the data, where the quiescent fraction of galaxies between the model and the data is well matched. In future studies it would be interesting to compare the \citet{peng10} model for the upper 5\% in environmental density to the cluster data, which would be a closer match to their density.

The UltraVISTA field is expected to contain a range of environmental densities. The measured SMF from these data should therefore be matched to a combination of the D1-D4 models. However, the right panel of Fig. \ref{fig:smf_peng_model} shows that even the lowest environmental density quartile D1 overpredicts the quiescent fraction of galaxies in the field of UltraVISTA at $z=1$. The caveat is that the separation of star-forming and quiescent galaxies is done differently between the data and the model. Our sample of UVJ-selected star-forming galaxies includes star-forming galaxies that are seen edge-on and therefore reddened by dust, whereas a rest-frame (U-B) colour selection, as applied in \citet{peng10}, identifies these objects as being on the red sequence. 

To reconcile the apparent disagreement between the data and the \citet{peng10} model in predicting the quiescent fraction of galaxies, we consider the following simplified analytical model where we only assume mass quenching and environmental quenching, and no additional merging. We apply this simplified model, based on the same principles as \citet{peng10}, to the GCLASS cluster data, but use the Schechter fits to the UltraVISTA data as a starting point. UltraVISTA is the limiting case where the dominant quenching process is mass-quenching.

We fit the cluster data by a combination of three functions that describe populations of star-forming galaxies, mass-quenched quiescent galaxies and environmentally-quenched quiescent galaxies. Two of these functions are given by the Schechter fits that were measured for the UltraVISTA field population. The quiescent population of UltraVISTA is expected to be primarily mass-quenched at the stellar mass range we study, so we take the shape of this SMF for the mass-quenched population and allow the normalisation to be adjusted by the fit. The SMF for star-forming galaxies is also taken from UltraVISTA, and since the functional form of this distribution is assumed to be independent of environment, we use the shape of this SMF and allow for a change in the normalisation. The third SMF, that describes the population of environmentally-quenched galaxies, is assumed to have the same shape as the SMF of star-forming galaxies, but the normalisation can be adapted in the fit. The sum of the functions for both quenched populations is fitted to the data points that describe the SMF for quiescent galaxies.

Now that the functional forms of the three populations that we fit are defined, the normalisations are adapted by fitting two free parameters in the following way. One free parameter $x$ describes how the three functions move relative to each other, and constrains the percentage of star-forming galaxies that is environmentally quenched by the cluster. This gives rise to a population of environmentally-quenched galaxies with a normalisation of $x$ compared to the star-forming galaxies. The star-forming galaxies are reduced by a factor of $(1-x)$. We do not change the relative amount of mass-quenched galaxies.  
The second free parameter describes the total normalisation of these three functions and has no direct meaning because the cluster and field are arbitrarily normalised with respect to each other.

\begin{figure}[ht]
\resizebox{\hsize}{!}{\includegraphics{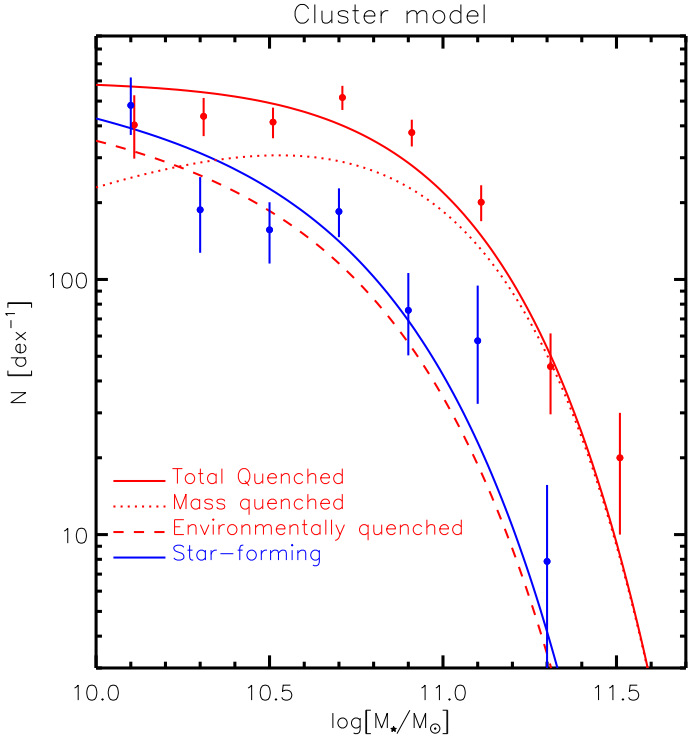}}
\caption{The cluster data, which were already given in Table~\ref{table:datapoints}, with the predictions from a simple quenching model based on \citet{peng10} in its most basic form, using UltraVISTA as a starting point. Mass quenching and environmental quenching are assumed to act independently. The blue Schechter function is the best fit to the star-forming galaxy population and the red solid line gives the best fit to the quiescent population. The red line is composed of a mass quenched population (dotted red line), and an environmentally quenched population (dashed line). This model needs 45\% additional environmental quenching compared to the field to yield the best fit to the data. Note that the red points have been offset by 0.01 dex for better visibility.}
\label{fig:smf_Q_model}
\end{figure}

We perform a maximum-likelihood fit to the data points for the star-forming and quiescent galaxies simultaneously, where we adapt these two parameters, and find a best fitting value of $x=0.45^{+0.04}_{-0.03}$. Assuming this simple picture we therefore find that, besides the quenching processes that also happen in the field, the cluster environment has to quench an additional 45\% of the galaxies to yield the best fit. In the Peng D4 model this environmental quenching fraction ranges from 0.17 at $10^{10}\,\rm{M_{\odot}}$ to 0.22 at $10^{11}\,\rm{M_{\odot}}$. 
Fig.~\ref{fig:smf_Q_model} shows the best fit to the observed SMF in the cluster based on this simple model. The blue and red solid lines give the simultaneous best fit to both galaxy types, for the star-forming and quiescent populations respectively. The red line is composed of a mass-quenched population (dotted red line), and an environmentally-quenched population (dashed line). The quiescent population at high ($>10^{10.2}\,\rm{M_{\odot}}$) masses is dominated by mass-quenched galaxies, while the population at lower stellar masses is dominated by environmentally-quenched galaxies. 

The best-fitting model does not yield a perfect representation of the data, since the model significantly overpredicts the number of quiescent galaxies in the low mass regime ($<10^{10.6}\,\rm{M_{\odot}}$). 
At intermediate masses around $10^{10.9}\,\rm{M_{\odot}}$ the model predicts about 30\% less galaxies than the data show. The overall Goodness-of-Fit for this model is 2.2 per degree of freedom. \citet{peng10} acknowledge that another term, due to merger quenching, is required to fit the data in SDSS and zCOSMOS. 

We know that mergers occur in clusters \citep[e.g.][]{vandokkum99}, and that the intra-cluster light builds up over time, probably by disruptions of relatively low mass galaxies \citep{martel12}. Also we know that BCGs have to grow in stellar mass over time \citep[e.g.][]{lidman12}, likely by consuming infalling galaxies. It is possible that these merging processes are required to reconcile the disagreement between the data and this model. 

The intriguing similarity in the shape of the quiescent SMF between the cluster and field environments at $z\sim 1$ suggests that there might be a simpler explanation than the \citet{peng10} model that does not involve a large amount of mergers. A similar internally-driven quenching mechanism might be responsible for the build-up quiescent population in both environments. 
We know that the age of a quiescent galaxy at a given stellar mass does not significantly depend on its environment \citep{thomas10,muzzin12}. However, for galaxies at a given stellar mass, their underlying dark matter (sub-)haloes at the time of formation might be different between the clusters and the field. "Environment quenching" could therefore refer to an internally driven process that is accelerated in cluster sub-haloes compared to the field. The finding that the cluster environment has already formed a large stellar mass content by $z\sim 1$ (see Fig.~\ref{fig:smf_normalised}) compared to the field, and that the fraction of quiescent galaxies is higher in the cluster than in the field, could be caused by a different evolution of the underlying dark matter haloes. 

A detailed study of the evolution of the (sub-)halo mass function, compared between cluster and field, is required to look into the different quenching scenarios. It is required to trace back the (sub-)haloes that host the galaxies we study to investigate how their progenitors merge and accrete during their formation history. Such a study could be useful to better understand the process of environmental quenching.

\section{Summary and Conclusions}\label{sec:conclusions}
In this paper we measured and compared the galaxy SMF at $z \sim 1$ in the high-density environments probed by GCLASS and the field environment from UltraVISTA. The GCLASS sample is composed of 10 rich, red-sequence selected clusters in the redshift range $0.86<z<1.34$. The $\rm{K_s}$-band selected catalogue based on observations in 11 photometric filters allowed us to estimate photometric redshifts and stellar masses for galaxies in the studied redshift range down to a stellar mass of $10^{10}\,\rm{M_\odot}$. The extensive spectroscopic sample of GCLASS covers the majority of the cluster members, and is critical to account for contaminants in the sample for which we only have photometric redshifts. Galaxies were separated by type (star-forming versus quiescent) based on their rest-frame U-V and V-J colours. For each galaxy in the photometric sample we estimated the probability that it is part of the cluster based on its type, stellar mass and clustercentric distance. This resulted in a statistically complete sample of cluster members to measure the SMF from. 

As a reference field SMF we used UltraVISTA, which is a new NIR survey that overlaps with COSMOS, resulting in 30 band photometric coverage over 1.62 square degrees. Analogously to GCLASS, sources were selected from the $\rm{K_s}$-band, and galaxies were separated between the star-forming and quiescent type using the rest-frame UVJ fluxes. This led to a measurement of the SMF for field galaxies at $0.85<z<1.20$ that is complete down to stellar masses of $10^{10}\,\rm{M_\odot}$. 

Under the assumption of a single Schechter function fit, we found that the shape of the SMF for star-forming galaxies is similar between the cluster and field environment, and that the combination of best-fitting Schechter parameters $\alpha$ and $M^{*}$ agree to 1$\sigma$ between the cluster and field. Furthermore, for the samples of quiescent galaxies we obtain a similar result. The shape of the SMF for quiescent galaxies is similar between the cluster and field at $\rm{M_{\star}} > 10^{10}\,\rm{M_\odot}$. The shape of the SMF for the total galaxy population is significantly different between the cluster and field. This is caused by a different fraction of quiescent galaxies in both environments. We find that there is a relative deficit of galaxies with low stellar masses in the cluster compared to the field. However, when we normalise the SMF by the total amount of matter in each respective part of the Universe, we find that there is a strong excess of galaxies over the entire stellar mass range we probe. This indicates that the cluster environment
must have been substantially more efficient in transforming mass into stars compared to the field. Note that this does not imply that field galaxies are less efficient, but rather it is the consequence of the fact that voids contain dark matter, but relatively few stars.

The similarity in the shape of the quiescent and star-forming SMF between the cluster and the field indicates that, if different processes are to be responsible for the quenching of star formation in different environments, these processes have to work in such a way that the shapes of the quiescent and star-forming SMF are similar in these different environments at $z=1$. This poses a challenge to analytical models that attempt to explain the build-up of the quiescent population by a combination of mass quenching and environment quenching. A simple model suggests that $45^{+4}_{-3}\%$ of the star-forming galaxies which normally would be forming stars in the field, would be quenched by the cluster. Although the physical processes that cause galaxies to quench environmentally are not yet completely understood, it is clear that a process like environmental quenching at $z \sim 1$ is important.

\begin{acknowledgements}
We thank Gregory Rudnick, Marijn Franx and Simone Weinmann for discussions and helpful suggestions for this study. Further we thank Jean-Charles Cuillandre for providing information on the MegaCam amplifier drift problem, and Michael Balogh for general feedback on the paper draft. 

R.F.J. van der Burg and H. Hoekstra acknowledge support from the Netherlands Organisation for Scientic Research grant number 639.042.814. C. Lidman is the recipient of an Australian Research Council Future Fellowship (program number FT0992259). H. Hildebrandt is supported by the Marie Curie IOF 252760, a CITA National Fellowship, and the DFG grant Hi 1495/2-1. R.D. gratefully acknowledges the support provided by the BASAL Center for Astrophysics and Associated Technologies (CATA), and by FONDECYT grant N. 1130528.

Based on observations obtained at the Gemini Observatory, which is operated by the Association of Universities for Research in Astronomy, Inc., under a cooperative agreement with the NSF on behalf of the Gemini partnership: the National Science Foundation (United States), the National Research Council (Canada), CONICYT (Chile), the Australian Research Council (Australia), Minist\'{e}rio da Ci\^{e}ncia, Tecnologia e Inova\c{c}\~{a}o (Brazil) and Ministerio de Ciencia, Tecnolog\'{i}a e Innovaci\'{o}n Productiva (Argentina).    
Based on observations obtained with MegaPrime/MegaCam, a joint project of CFHT and CEA/DAPNIA, at the Canada-France-Hawaii Telescope (CFHT) which is operated by the National Research Council (NRC) of Canada, the Institute National des Sciences de l'Univers of the Centre National de la Recherche Scientifique of France, and the University of Hawaii. 
Based on observations obtained with WIRCam, a joint project of CFHT, Taiwan, Korea, Canada, France, and the Canada-France-Hawaii Telescope (CFHT) which is operated by the National Research Council (NRC) of Canada, the Institute National des Sciences de l'Univers of the Centre National de la Recherche Scientifique of France, and the University of Hawaii. 
This work is based in part on observations made with the Spitzer Space Telescope, which is operated by the Jet Propulsion Laboratory, California Institute of Technology under a contract with NASA. Support for this work was provided by NASA through an award issued by JPL/Caltech.
This paper includes data gathered with the 6.5 meter Magellan Telescopes located at Las Campanas Observatory, Chile.  
This work is based on observations obtained at the CTIO Blanco 4-m telescopes, which are operated by the Association of Universities for Research in Astronomy Inc. (AURA), under a cooperative agreement with the NSF as part of the National Optical Astronomy Observatories (NOAO). 
Based on observations that were carried out using the Very Large Telescope at the ESO Paranal Observatory. 
Based on data products from observations made with ESO Telescopes at the La Silla Paranal Observatories under ESO programme ID 179.A-2005 and on data products produced by TERAPIX and the Cambridge Astronomy Survey Unit on behalf of the UltraVISTA consortium.
\end{acknowledgements}

\bibliographystyle{aa} 
\bibliography{MasterRefs} 
\begin{appendix}
\section{Data processing and catalogue creation}\label{catalogcreation}
This Appendix is meant to give a more elaborate description of the data reduction steps (Sect. \ref{app:datareduction}) and in particular the procedure for homogenising the PSF and measuring colours using Gaussian weighted apertures (Sect. \ref{app:gausspsf}). Because we combine photometric data over a wide range of wavelengths and for clusters that are both in the Northern and Southern sky, we necessarily have to combine data from different telescopes and/or instruments.

\subsection{Photometric data reduction}\label{app:datareduction}
The standard reduction steps include bias and flatfield corrections. Although the images have been flatfielded (e.g. by $Elixir$ for the MegaCam data) to yield a uniform zeropoint for the source fluxes, there are still residual background patterns due to scattered light, fringe residuals, and amplifier drift (Cuillandre, private communication). These patterns are reasonably stable over time, and since most exposures in a given filter have been taken consecutively on the same night, we can subtract these background effects. We do this by using the dithered pattern of observations to differentiate signals that are on a fixed position on the ccd array from sky-bound signals.

To remove cosmic rays from ccd images one usually compares different frames of the same part of the sky. However, since we only have a few deep exposures in some of the filters, the number of overlapping frames of our data set is not always sufficient to be able to identify all cosmic rays. Therefore we remove cosmic rays by using the Laplacian Cosmic Ray Identification method \citep{dokkumcosmics}, which works on individual images. We optimise the parameters in the setup of the code such that we minimise the amount of false positives (bright stars) and false negatives. We do this by testing the code on a range of images with different seeing. The only parameter that has a significant influence on the fraction of false positives and false negatives is $objlim$, which we take to be 3.0.

We obtain astrometric and relative photometric solutions for each chip using $\tt{SCAMP}$ \citep{scamp}, where we use all exposures in a given filter for all clusters at once to effectively increase the source density, and obtain stable solutions. As a reference catalogue we use SDSS-DR7 data, or the USNO-B catalogue whenever a cluster falls outside the SDSS footprint. This leads to consistent astrometric solutions between the different filters with an internal scatter of about $0.05''$.

\begin{table*}   
\caption{The GCLASS photometric data set. The instruments used for the different clusters and filters are indicated. The limiting magnitudes reported are median 5-$\sigma$ flux measurement limits for point sources measured with a Gaussian weight function.}
\label{table:data} 
\begin{center}
\centering 
\begin{tabular}{c c c c c c c c c c c c c c} 
\hline\hline 
Name$^{\mathrm{a}}$ & $u_{\mathrm{lim}}$ &$g_{\mathrm{lim}}$ &$r_{\mathrm{lim}}$ &$i_{\mathrm{lim}}$ &$z_{\mathrm{lim}}$ &$\mathrm{J_{lim}}$ &$\mathrm{K_{s,lim}}$&$\mathrm{3.6\mu m_{lim}}$&$\mathrm{4.5\mu m_{lim}}$&$\mathrm{5.8\mu m_{lim}}$&$\mathrm{8.0\mu m_{lim}}$ \\ 
&[$\mathrm{mag_{AB}}$]&[$\mathrm{mag_{AB}}$]&[$\mathrm{mag_{AB}}$]&[$\mathrm{mag_{AB}}$]&[$\mathrm{mag_{AB}}$]&[$\mathrm{mag_{AB}}$]
&[$\mathrm{mag_{AB}}$]&[$\mathrm{mag_{AB}}$]&[$\mathrm{mag_{AB}}$]&[$\mathrm{mag_{AB}}$]&[$\mathrm{mag_{AB}}$]\\
\hline 
SpARCS-0034 &23.1$^{\mathrm{b}}$& 25.3$^{\mathrm{b}}$& 24.4$^{\mathrm{b}}$& 24.3$^{\mathrm{b}}$& 23.9$^{\mathrm{c}}$& 22.5$^{\mathrm{e}}$& 22.2$^{\mathrm{e}}$& 21.4$^{\mathrm{g}}$& 21.2$^{\mathrm{g}}$& 19.7$^{\mathrm{g}}$& 19.6$^{\mathrm{g}}$\\
SpARCS-0035 &24.4$^{\mathrm{b}}$& 25.4$^{\mathrm{b}}$& 24.9$^{\mathrm{b}}$& 24.3$^{\mathrm{b}}$& 23.6$^{\mathrm{c}}$& 24.1$^{\mathrm{f}}$& 23.4$^{\mathrm{f}}$& 22.8$^{\mathrm{g}}$& 22.3$^{\mathrm{g}}$& 20.8$^{\mathrm{g}}$& 20.4$^{\mathrm{g}}$\\
SpARCS-0036 &22.9$^{\mathrm{b}}$& 25.1$^{\mathrm{b}}$& 24.4$^{\mathrm{b}}$& 23.7$^{\mathrm{b}}$& 23.5$^{\mathrm{c}}$& 22.7$^{\mathrm{e}}$& 21.5$^{\mathrm{e}}$& 21.2$^{\mathrm{g}}$& 21.1$^{\mathrm{g}}$& 19.9$^{\mathrm{g}}$& 19.6$^{\mathrm{g}}$\\
SpARCS-0215 &24.8$^{\mathrm{a}}$& 25.1$^{\mathrm{b}}$& 24.7$^{\mathrm{b}}$& 24.4$^{\mathrm{b}}$& 23.7$^{\mathrm{a}}$& 22.8$^{\mathrm{e}}$& 22.0$^{\mathrm{e}}$& 21.3$^{\mathrm{g}}$& 21.1$^{\mathrm{g}}$& 19.5$^{\mathrm{g}}$& 19.4$^{\mathrm{g}}$\\
SpARCS-1047 &25.5$^{\mathrm{a}}$& 25.7$^{\mathrm{a}}$& 25.0$^{\mathrm{a}}$& 24.7$^{\mathrm{a}}$& 23.8$^{\mathrm{a}}$& 23.1$^{\mathrm{d}}$& 22.3$^{\mathrm{d}}$& 21.6$^{\mathrm{g}}$& 21.3$^{\mathrm{g}}$& 19.7$^{\mathrm{g}}$& 19.7$^{\mathrm{g}}$\\
SpARCS-1051 &25.6$^{\mathrm{a}}$& 25.8$^{\mathrm{a}}$& 25.2$^{\mathrm{a}}$& 25.0$^{\mathrm{a}}$& 24.0$^{\mathrm{a}}$& 23.2$^{\mathrm{d}}$& 22.4$^{\mathrm{d}}$& 21.7$^{\mathrm{g}}$& 21.3$^{\mathrm{g}}$& 19.8$^{\mathrm{g}}$& 19.7$^{\mathrm{g}}$\\
SpARCS-1613 &25.5$^{\mathrm{a}}$& 26.0$^{\mathrm{a}}$& 25.4$^{\mathrm{a}}$& 24.7$^{\mathrm{a}}$& 24.0$^{\mathrm{a}}$& 23.1$^{\mathrm{d}}$& 22.7$^{\mathrm{d}}$& 22.7$^{\mathrm{g}}$& 22.6$^{\mathrm{g}}$& 21.2$^{\mathrm{g}}$& 20.9$^{\mathrm{g}}$\\
SpARCS-1616 &25.1$^{\mathrm{a}}$& 25.7$^{\mathrm{a}}$& 25.1$^{\mathrm{a}}$& 24.8$^{\mathrm{a}}$& 23.5$^{\mathrm{a}}$& 23.3$^{\mathrm{d}}$& 22.7$^{\mathrm{d}}$& 22.6$^{\mathrm{g}}$& 22.4$^{\mathrm{g}}$& 21.2$^{\mathrm{g}}$& 20.9$^{\mathrm{g}}$\\
SpARCS-1634 &25.6$^{\mathrm{a}}$& 26.1$^{\mathrm{a}}$& 25.6$^{\mathrm{a}}$& 25.1$^{\mathrm{a}}$& 24.4$^{\mathrm{a}}$& 23.7$^{\mathrm{d}}$& 23.1$^{\mathrm{d}}$& 23.2$^{\mathrm{g}}$& 23.2$^{\mathrm{g}}$& 21.6$^{\mathrm{g}}$& 21.3$^{\mathrm{g}}$\\
SpARCS-1638 &25.4$^{\mathrm{a}}$& 25.9$^{\mathrm{a}}$& 25.4$^{\mathrm{a}}$& 25.1$^{\mathrm{a}}$& 24.2$^{\mathrm{a}}$& 23.4$^{\mathrm{d}}$& 22.8$^{\mathrm{d}}$& 23.0$^{\mathrm{g}}$& 23.1$^{\mathrm{g}}$& 21.3$^{\mathrm{g}}$& 21.3$^{\mathrm{g}}$\\
\hline
\end{tabular}
\end{center}
\begin{list}{}{}
\item[$^{\mathrm{a}}$] MegaCam, Canada-France-Hawaii Telescope (CFHT)
\item[$^{\mathrm{b}}$] IMACS, Magellan Telescope
\item[$^{\mathrm{c}}$] MOSAIC-II, Blanco Telescope, Cerro Tololo Inter-American Observatory (CTIO)
\item[$^{\mathrm{d}}$] WIRCam, Canada-France-Hawaii Telescope (CFHT)
\item[$^{\mathrm{e}}$] ISPI, Blanco Telescope, Cerro Tololo Inter-American Observatory (CTIO)
\item[$^{\mathrm{f}}$] HAWK-I, Very Large Telescope (VLT) UT4
\item[$^{\mathrm{g}}$] IRAC, Spitzer Space Telescope
\end{list}
\end{table*}

\subsection{PSF homogenisation and colour measurements}\label{app:gausspsf}
Because the shape and size of the point spread function (PSF) are different between exposures and filters, it is non-trivial to measure accurate colours of a galaxy. The simplest approach would be to take the ratio of the total flux of a galaxy in different bands, but this requires very large photometric apertures: for background-limited observations these are very noisy.

However, a reliable colour measurement for the purpose of photometric redshift determination can also be made by taking the ratio of aperture fluxes in different bands, provided these apertures represent the same intrinsic part of the galaxy. We have followed this approach here, based on a modification of the Gaussian-aperture-and-PSF (GaaP) photometry method \citep{kuijken08}.

The first step is to convolve each image with a suitable position-dependent kernel that modifies its PSF into a uniform size, circular and Gaussian. This kernel can be constructed using the shapelet \citep{refregier03,kuijken06} formalism, as was done in the \textit{local} approach described in \citet{hildebrandt12}, with one modification: here we allow the resulting PSF size for each image to be different. Specifically, for each filter and field we choose the size of the resulting PSF to be slightly larger (ca. 10\%) than the median gaussian radius of all bright stars in the images. To obtain a stable PSF in the stacked images for each filter we Gaussianise the PSF of the individual astrometrically corrected exposures before stacking.

Following \citet{kuijken08} we then measure fluxes in the following way. Instead of using a function where the weight is either 0 or 1, as is the case for regular aperture photometry measured with a top-hat weighting function (e.g. by running $\tt{SExtractor}$ in dual image mode), we use a smooth weight function that makes use of the fact that the S/N for each pixel decreases away from the peak pixel. When the PSF in each filter follows a Gaussian profile, the choice to perform photometry using a Gaussian weight function is computationally convenient, as we show next.

\citet{kuijken08} defines the "Gaussian-aperture-and-PSF" flux $F_q$ as the Gaussian weighted flux a source would have if it were observed with a Gaussian PSF with the same width $q$ as the weight function. Hence
\begin{equation}\label{eq:fq1}
F_{q} \equiv \int \mathrm{d} \mathbf{r} \, \mathrm{e}^{-r^2/2q^2} \int \mathrm{d}\mathbf{r}'\,S(\mathbf{r}') \,\frac{\mathrm{e}^{-(\mathbf{r-r'})^2/2q^2}}{2\pi q^2}, 
\end{equation}
where $S$ is the intrinsic light distribution of the source (i.e. before smearing with the PSF) and $q$ is the scale radius of the weight function. 
It is straightforward to simplify Eq. \ref{eq:fq1} to
\begin{equation}
F_{q} = \int \mathrm{d} \mathbf{r} \, \frac{1}{2} \,S(\mathbf{r}) \,\mathrm{e}^{-r^2/4q^2},
\end{equation}
which shows that $F_{q}$ is a Gaussian-aperture photometric measurement of the \textit{intrinsic} galaxy. 

After gaussianising the images, $S$ has already been convolved with a Gaussian that has a constant dispersion $g_{PSF}$ for each stacked image. The flux distribution on the ccd is therefore
\begin{equation}
I(\mathbf{r}) = \int \mathrm{d} \mathbf{r'} \, S(\mathbf{r'}) \,\frac{\mathrm{e}^{-(\mathbf{r-r'})^2/2\,g_{PSF}^2}}{2\pi \,g_{PSF}^2 }.
\end{equation}

Analytically we have an identical expression for $F_{q}$
\begin{equation}\label{eq:fq2}
F_{q} = \int \mathrm{d}\mathbf{r}' I(\mathbf{r}') \, \frac{q^2}{2q^2-g_{PSF}^2} \,\mathrm{e}^{-(\mathbf{r-r'})^2/2(2q^2-g_{PSF}^2)},
\end{equation}
which thus shows that the same intrinsic aperture flux $F_{q}$ can be measured from images with a range of Gaussian PSF sizes. 
Therefore, from our PSF-gaussianised images, we can measure colours of the same intrinsic part of the galaxy if we use Gaussian weight functions to measure fluxes. Note that it is no longer necessary that the stacks of the different filters have a PSF with the same Gaussian FWHM, as long as the weight function is adapted accordingly for each filter. 

We adjust $q$ to ensure the aperture roughly matches each galaxy's size, to optimise the S/N. We base our choice for $q$ on the SExtractor parameter FLUX\_RADIUS measured in the $\rm{K_{s}}$-band image, such that $q=0.85\cdot$FLUX\_RADIUS. The factor of 0.85 is chosen to optimise the S/N of a source with a circular Gaussian PSF-profile. Further we make sure that $q$ is chosen such that $q > g_{PSF}$ in all filters.

This method is applied to measure fluxes in the $u-\rm{K_s}$-bands, but since the IRAC data suffer from a much larger PSF we work in a two stage process to incorporate the IRAC fluxes in a way that reduces the problems from confusion. We construct a 2-stage multi-colour catalogue where we multiply the IRAC flux measured in the bigger aperture with the fractional difference of the $\rm{K_s}$-band flux measured in the small and bigger aperture. This way we effectively correct the IRAC flux for blending with nearby objects by assuming these neighbours have the same ($\rm{K_s}$-IRAC) colour as the source. For contaminating galaxies this is often the case. To verify that any residual blending in the IRAC bands does not affect our results, we repeated the analysis while excluding the IRAC data in all SED fits. We find no bias in the stellar mass estimates, and even for the lowest masses ($\rm{M}_{\star}=10^{10}\rm{M}_{\odot}$), $68\%$ of the estimated stellar masses differ by less than 0.05 dex from our fiducial analysis.
 
We calibrate the photometric zeropoints on a catalogue basis by making use of the universality of the stellar locus \citep{slr}. We use stellar data from \citet{covey07}, containing 600,000 point sources selected from the SDSS and 2MASS surveys. By applying linear colour terms we compare these colours to stars measured with the filter sets in the telescope we used. Note that these data are especially favourable to calibrate the zeropoints using the stellar locus since the amount of galactic dust is very low in these fields. We adapt the zeropoints of the $ugriz\rm{JK_{s}}$ filters to bring the colours of stars in our data in line with the reference catalogue. Corrections are typically on the order of 0.05 magnitudes. To account for uncertainties in the absolute zeropoint of IRAC, we included a 10$\%$ systematic error to the IRAC fluxes.

After gaussianisation, the background noise in the images is correlated between pixels. Therefore we estimate the errors on the flux measurements in the stacks of each filter by measuring the fluctuations in the flux values measured from apertures that are randomly placed on the images. We take account of the non-uniform exposure time over the image stacks. Table \ref{table:data} shows an overview of the median 5-$\sigma$ flux measurements for point sources in each filter and each cluster.

\end{appendix}

\begin{appendix}
\section{Field SMF measurements from GCLASS}\label{app:fieldfield}
\begin{figure*}
\resizebox{\hsize}{!}{\includegraphics{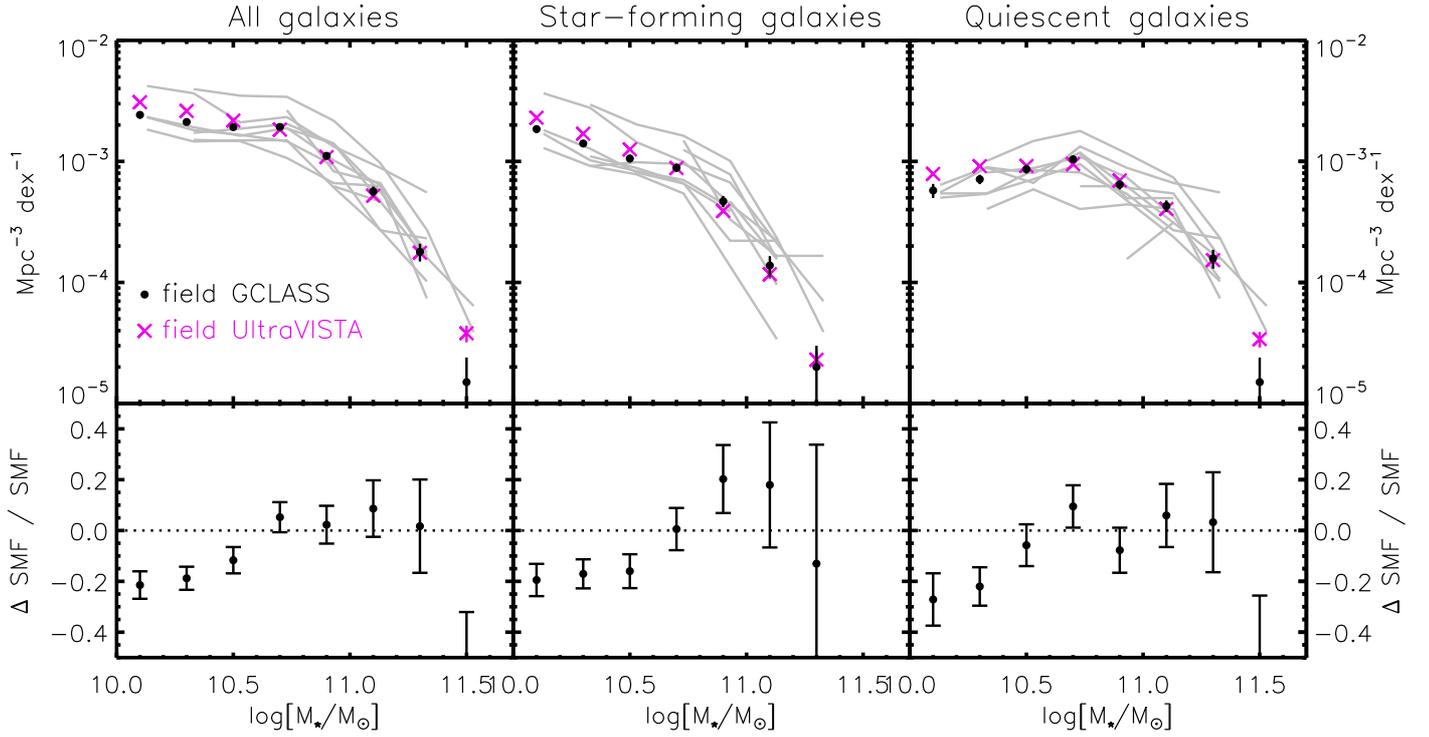}}
\caption{The UltraVISTA (magenta) versus GCLASS (black) field measurements. Left panel: the total galaxy population in both fields. Middle panel: the SMF for the subset of star-forming galaxies. Right panel: the subset of quiescent galaxies. Error bars show the $68\%$ confidence regions for Poisson error bars. The grey curves show the 10 contributions to the field SMF around the GCLASS clusters, which differ because of cosmic variance due to the small volumes probed in these individual fields. Also the fields contribute only down to a particular mass respecting the varying depths of the GCLASS fields. Bottom panels: the fractional differences between the two field measurements, given by $\frac{\rm{GCLASS-UltraVISTA}}{\rm{UltraVISTA}}$, together with the estimated errors.}
\label{fig:field_versus_field}
\end{figure*}

Thanks to the relatively wide areas that were observed to obtain the GCLASS multi-colour catalogues ($15'\times 15'$ centred on the clusters in the Northern sky, and $10' \times 10'$ for the clusters in the Southern sky), these data can also be used to study galaxies outside the clusters and hence to measure the SMF of the general field. In this appendix we measure the field SMF from GCLASS in the redshift range $0.85<z<1.20$ and compare this to the field SMF measured from UltraVISTA. 

Since the UltraVISTA sample is based on a relatively deep (compared to GCLASS) 30-band photometric catalogue, it is complete in the mass range ($\rm{M_{\star}} > 10^{10}\, \rm{M_\odot}$) at this redshift range. A comparison between GCLASS and UltraVISTA may reveal possible systematic differences in the stellar mass catalogues, and any residual incompleteness in GCLASS.

To minimise the contamination by cluster galaxies in the sample, we use a conservative selection of field galaxies in GCLASS. A galaxy is considered as part of the field when it is separated from the cluster centre by more than the angular distance that corresponds to 1.5 Mpc at the redshift of the cluster. Furthermore we require a field galaxy to have a photometric redshift $|z_{\mathrm{phot}}-z_{\mathrm{cluster}}|>0.05$. 
After taking account of the areas masked by bright stars, this results in a total probed volume of the field that is $\sim 6$ times smaller in GCLASS compared to UltraVISTA. Since the 10 GCLASS pointings have different depths, we have to take account of the estimated mass-completeness of the detection bands. This is measured similarly as Sect.~\ref{sec:stellarmasses}, but using a redshift limit of 1.20 in each field (instead of the individual cluster redshifts). This way we correct for Malmquist bias in a similar way as in the $1/V_{\rm{max}}$ weighting method.

Fig.~\ref{fig:field_versus_field} shows a comparison of the field SMF measured in the GCLASS (black) and UltraVISTA (magenta) surveys. The curves are normalised with respect to the total volume subtended by these surveys. The grey curves show the contributions to the field SMF of the 10 individual GCLASS fields. These contributions differ between the pointings because their depths are different, and also the area surrounding the cluster that is part of the field differs. The differences in the grey curves are further caused by cosmic variance. The field in the SpARCS-1047 image for example is significantly overdense in the redshift bin $0.85<z<1.20$. Note however that, when these 10 independent sight-lines of GCLASS are combined, the uncertainty due to cosmic variance is greatly reduced \citep{somerville04}. 

There is generally a good agreement between the field SMF measurements from GCLASS and UltraVISTA, especially at the high-mass end. This indicates that there are no substantial systematic differences between the two catalogues this study is based on. At the low-mass end of the SMF there are some systematic differences in both the star-forming and quiescent population, increasing to several $\sim10\%$ in the lowest mass bins. In Sect.~\ref{sec:clustersmf} we explained that we corrected the GCLASS cluster SMF data by these completeness correction factors. That way we can not only compare the cluster and field qualitatively, but have a more realistic view on the absolute Schechter parameters. Note that this additional completeness correction does not change any of the qualitative statements in this paper, nor affects the conclusions of this paper in any way.

\end{appendix}

\end{document}